\documentclass[longauth]{aa} 

\bibpunct{(}{)}{;}{a}{}{,} 
\usepackage{graphicx}
\usepackage{booktabs} 
\usepackage{siunitx}

\usepackage{txfonts}
\usepackage[normalem]{ulem}
\begin{document}

\title{
{LMC N132D: {A} mature supernova remnant with a power-law gamma-ray spectrum extending beyond 8 TeV}
}

\titlerunning{N132D: Mature but luminous SNR with a power-law spectrum}
\authorrunning{H.E.S.S. Collaboration}
\author{\fontsize{8.8}{10.4}\selectfont{H.E.S.S. Collaboration
\and H.~Abdalla \inst{\ref{UNAM}}
\and F.~Aharonian \inst{\ref{DIAS},\ref{MPIK},\ref{RAU}}
\and F.~Ait~Benkhali \inst{\ref{MPIK}}
\and E.O.~Ang\"uner \inst{\ref{CPPM}}
\and C.~Arcaro \inst{\ref{NWU}}
\and C.~Armand \inst{\ref{LAPP}}
\and T.~Armstrong \inst{\ref{Oxford}}
\and H.~Ashkar \inst{\ref{IRFU}}
\and M.~Backes \inst{\ref{UNAM},\ref{NWU}}
\and V.~Baghmanyan \inst{\ref{IFJPAN}}
\and V.~Barbosa~Martins \inst{\ref{DESY}}
\and A.~Barnacka \inst{\ref{UJK}}
\and M.~Barnard \inst{\ref{NWU}}
\and R.~Batzofin \inst{\ref{WITS}}
\and Y.~Becherini \inst{\ref{Linnaeus}}
\and D.~Berge \inst{\ref{DESY}}
\and K.~Bernl\"ohr \inst{\ref{MPIK}}
\and B.~Bi \inst{\ref{IAAT}}
\and M.~B\"ottcher \inst{\ref{NWU}}
\and C.~Boisson \inst{\ref{LUTH}}
\and J.~Bolmont \inst{\ref{LPNHE}}
\and M.~de~Bony~de~Lavergne \inst{\ref{LAPP}}
\and M.~Breuhaus \inst{\ref{MPIK}}
\and R.~Brose \inst{\ref{DIAS}}
\and F.~Brun \inst{\ref{IRFU}}
\and T.~Bulik \inst{\ref{UWarsaw}}
\and T.~Bylund \inst{\ref{Linnaeus}}
\and F.~Cangemi \inst{\ref{LPNHE}}
\and S.~Caroff \inst{\ref{LPNHE}}
\and S.~Casanova \inst{\ref{IFJPAN}}
\and J.~Catalano \inst{\ref{ECAP}}
\and P.~Chambery \inst{\ref{CENBG}}
\and T.~Chand \inst{\ref{NWU}}
\and A.~Chen \inst{\ref{WITS}}
\and G.~Cotter \inst{\ref{Oxford}}
\and M.~Cury{\l}o \inst{\ref{UWarsaw}}
\and J.~Damascene~Mbarubucyeye \inst{\ref{DESY}}
\and I.D.~Davids \inst{\ref{UNAM}}
\and J.~Davies \inst{\ref{Oxford}}
\and J.~Devin \inst{\ref{CENBG}}
\and A.~Djannati-Ata\"i \inst{\ref{APC}}
\and A.~Dmytriiev \inst{\ref{LUTH}}
\and A.~Donath \inst{\ref{MPIK}}
\and V.~Doroshenko \inst{\ref{IAAT}}
\and L.~Dreyer \inst{\ref{NWU}}
\and L.~Du~Plessis \inst{\ref{NWU}}
\and C.~Duffy \inst{\ref{Leicester}}
\and K.~Egberts \inst{\ref{UP}}
\and S.~Einecke \inst{\ref{Adelaide}}
\and J.-P.~Ernenwein \inst{\ref{CPPM}}
\and S.~Fegan \inst{\ref{LLR}}
\and K.~Feijen \inst{\ref{Adelaide}}
\and A.~Fiasson \inst{\ref{LAPP}}
\and G.~Fichet~de~Clairfontaine \inst{\ref{LUTH}}
\and G.~Fontaine \inst{\ref{LLR}}
\and F.~Lott \inst{\ref{UNAM}}
\and M.~F\"u{\ss}ling \inst{\ref{DESY}}
\and S.~Funk \inst{\ref{ECAP}}
\and S.~Gabici \inst{\ref{APC}}
\and Y.A.~Gallant \inst{\ref{LUPM}}
\and G.~Giavitto \inst{\ref{DESY}}
\and L.~Giunti \inst{\ref{APC},\ref{IRFU}}
\and D.~Glawion \inst{\ref{ECAP}}
\and J.F.~Glicenstein \inst{\ref{IRFU}}
\and M.-H.~Grondin \inst{\ref{CENBG}}
\and S.~Hattingh \inst{\ref{NWU}}
\and M.~Haupt \inst{\ref{DESY}}
\and G.~Hermann \inst{\ref{MPIK}}
\and J.A.~Hinton \inst{\ref{MPIK}}
\and W.~Hofmann \inst{\ref{MPIK}}
\and C.~Hoischen \inst{\ref{UP}}
\and T.~L.~Holch \inst{\ref{DESY}}
\and M.~Holler \inst{\ref{LFUI}}
\and M.~H\"{o}rbe \inst{\ref{Oxford}}
\and D.~Horns \inst{\ref{HH}}
\and Zhiqiu~Huang \inst{\ref{MPIK}}
\and D.~Huber \inst{\ref{LFUI}}
\and M.~Jamrozy \inst{\ref{UJK}}
\and F.~Jankowsky \inst{\ref{LSW}}
\and V.~Joshi \inst{\ref{ECAP}}
\and I.~Jung-Richardt \inst{\ref{ECAP}}
\and E.~Kasai \inst{\ref{UNAM}}
\and K.~Katarzy{\'n}ski \inst{\ref{NCUT}}
\and U.~Katz \inst{\ref{ECAP}}
\and D.~Khangulyan \inst{\ref{Rikkyo}}
\and B.~Kh\'elifi \inst{\ref{APC}}
\and S.~Klepser \inst{\ref{DESY}}
\and W.~Klu\'{z}niak \inst{\ref{NCAC}}
\and Nu.~Komin \inst{\ref{WITS}}\protect\footnotemark[1]
\and R.~Konno \inst{\ref{DESY}}
\and K.~Kosack \inst{\ref{IRFU}}
\and D.~Kostunin \inst{\ref{DESY}}
\and M.~Kreter \inst{\ref{NWU}}
\and G.~Kukec~Mezek \inst{\ref{Linnaeus}}
\and A.~Kundu \inst{\ref{NWU}}
\and G.~Lamanna \inst{\ref{LAPP}}
\and S.~Le Stum \inst{\ref{CPPM}}
\and A.~Lemi\`ere \inst{\ref{APC}}
\and M.~Lemoine-Goumard \inst{\ref{CENBG}}
\and J.-P.~Lenain \inst{\ref{LPNHE}}
\and F.~Leuschner \inst{\ref{IAAT}}
\and C.~Levy \inst{\ref{LPNHE}}
\and T.~Lohse \inst{\ref{HUB}}
\and A.~Luashvili \inst{\ref{LUTH}}
\and I.~Lypova \inst{\ref{LSW}}
\and J.~Mackey \inst{\ref{DIAS}}
\and J.~Majumdar \inst{\ref{DESY}}
\and D.~Malyshev \inst{\ref{ECAP}}
\and D.~Malyshev \inst{\ref{IAAT}}
\and V.~Marandon \inst{\ref{MPIK}}
\and P.~Marchegiani \inst{\ref{WITS}}
\and A.~Marcowith \inst{\ref{LUPM}}
\and A.~Mares \inst{\ref{CENBG}}
\and G.~Mart\'i-Devesa \inst{\ref{LFUI}}
\and R.~Marx \inst{\ref{LSW}}
\and G.~Maurin \inst{\ref{LAPP}}
\and P.J.~Meintjes \inst{\ref{UFS}}
\and M.~Meyer \inst{\ref{ECAP}}
\and A.~Mitchell \inst{\ref{MPIK}}
\and R.~Moderski \inst{\ref{NCAC}}
\and L.~Mohrmann \inst{\ref{ECAP}}
\and A.~Montanari \inst{\ref{IRFU}}
\and C.~Moore \inst{\ref{Leicester}}
\and E.~Moulin \inst{\ref{IRFU}}
\and J.~Muller \inst{\ref{LLR}}
\and T.~Murach \inst{\ref{DESY}}
\and K.~Nakashima \inst{\ref{ECAP}}
\and M.~de~Naurois \inst{\ref{LLR}}
\and A.~Nayerhoda \inst{\ref{IFJPAN}}
\and H.~Ndiyavala \inst{\ref{NWU}}
\and J.~Niemiec \inst{\ref{IFJPAN}}
\and A.~Priyana~Noel \inst{\ref{UJK}}
\and P.~O'Brien \inst{\ref{Leicester}}
\and L.~Oberholzer \inst{\ref{NWU}}
\and H.~Odaka \inst{\ref{Rikkyo}}
\and S.~Ohm \inst{\ref{DESY}}
\and L.~Olivera-Nieto \inst{\ref{MPIK}}
\and E.~de~Ona~Wilhelmi \inst{\ref{DESY}}
\and M.~Ostrowski \inst{\ref{UJK}}
\and S.~Panny \inst{\ref{LFUI}}
\and M.~Panter \inst{\ref{MPIK}}
\and R.D.~Parsons \inst{\ref{HUB}}
\and G.~Peron \inst{\ref{MPIK}}
\and S.~Pita \inst{\ref{APC}}
\and V.~Poireau \inst{\ref{LAPP}}
\and D.A.~Prokhorov \inst{\ref{GRAPPA}}\protect\footnotemark[1]
\and H.~Prokoph \inst{\ref{DESY}}
\and G.~P\"uhlhofer \inst{\ref{IAAT}}
\and M.~Punch \inst{\ref{APC},\ref{Linnaeus}}
\and A.~Quirrenbach \inst{\ref{LSW}}
\and P.~Reichherzer \inst{\ref{IRFU}}
\and A.~Reimer \inst{\ref{LFUI}}
\and O.~Reimer \inst{\ref{LFUI}}
\and Q.~Remy \inst{\ref{MPIK}}
\and M.~Renaud \inst{\ref{LUPM}}
\and B.~Reville \inst{\ref{MPIK}}
\and F.~Rieger \inst{\ref{MPIK}}
\and C.~Romoli \inst{\ref{MPIK}}
\and G.~Rowell \inst{\ref{Adelaide}}
\and B.~Rudak \inst{\ref{NCAC}}
\and H.~Rueda Ricarte \inst{\ref{IRFU}}
\and E.~Ruiz-Velasco \inst{\ref{MPIK}}
\and V.~Sahakian \inst{\ref{YPI}}
\and S.~Sailer \inst{\ref{MPIK}}
\and H.~Salzmann \inst{\ref{IAAT}}
\and D.A.~Sanchez \inst{\ref{LAPP}}
\and A.~Santangelo \inst{\ref{IAAT}}
\and M.~Sasaki \inst{\ref{ECAP}}
\and J.~Sch\"afer \inst{\ref{ECAP}}
\and F.~Sch\"ussler \inst{\ref{IRFU}}
\and H.M.~Schutte \inst{\ref{NWU}}
\and U.~Schwanke \inst{\ref{HUB}}
\and M.~Senniappan \inst{\ref{Linnaeus}}
\and A.S.~Seyffert \inst{\ref{NWU}}
\and J.N.S.~Shapopi \inst{\ref{UNAM}}
\and K.~Shiningayamwe \inst{\ref{UNAM}}
\and R.~Simoni \inst{\ref{GRAPPA}}\protect\footnotemark[1]
\and A.~Sinha \inst{\ref{LUPM}}
\and H.~Sol \inst{\ref{LUTH}}
\and A.~Specovius \inst{\ref{ECAP}}
\and S.~Spencer \inst{\ref{Oxford}}
\and M.~Spir-Jacob \inst{\ref{APC}}
\and {\L.}~Stawarz \inst{\ref{UJK}}
\and R.~Steenkamp \inst{\ref{UNAM}}
\and C.~Stegmann \inst{\ref{UP},\ref{DESY}}
\and S.~Steinmassl \inst{\ref{MPIK}}
\and C.~Steppa \inst{\ref{UP}}
\and L.~Sun \inst{\ref{GRAPPA}}
\and T.~Takahashi \inst{\ref{Rikkyo}}
\and T.~Tanaka \inst{\ref{Rikkyo}}
\and T.~Tavernier \inst{\ref{IRFU}}
\and A.M.~Taylor \inst{\ref{DESY}}
\and R.~Terrier \inst{\ref{APC}}
\and J.~H.E.~Thiersen \inst{\ref{NWU}}
\and C.~Thorpe-Morgan \inst{\ref{IAAT}}
\and M.~Tluczykont \inst{\ref{HH}}
\and L.~Tomankova \inst{\ref{ECAP}}
\and M.~Tsirou \inst{\ref{MPIK}}
\and N.~Tsuji \inst{\ref{Rikkyo}}
\and R.~Tuffs \inst{\ref{MPIK}}
\and Y.~Uchiyama \inst{\ref{Rikkyo}}
\and D.J.~van~der~Walt \inst{\ref{NWU}}
\and C.~van~Eldik \inst{\ref{ECAP}}
\and C.~van~Rensburg \inst{\ref{UNAM}}
\and B.~van~Soelen \inst{\ref{UFS}}
\and G.~Vasileiadis \inst{\ref{LUPM}}
\and J.~Veh \inst{\ref{ECAP}}
\and C.~Venter \inst{\ref{NWU}}
\and P.~Vincent \inst{\ref{LPNHE}}
\and J.~Vink \inst{\ref{GRAPPA}}\protect\footnotemark[1]
\and H.J.~V\"olk \inst{\ref{MPIK}}
\and S.J.~Wagner \inst{\ref{LSW}}
\and J.~Watson \inst{\ref{Oxford}}
\and F.~Werner \inst{\ref{MPIK}}
\and R.~White \inst{\ref{MPIK}}
\and A.~Wierzcholska \inst{\ref{IFJPAN}}
\and Yu~Wun~Wong \inst{\ref{ECAP}}
\and H.~Yassin \inst{\ref{NWU}}
\and A.~Yusafzai \inst{\ref{ECAP}}
\and M.~Zacharias \inst{\ref{LUTH}}
\and R.~Zanin \inst{\ref{MPIK}}
\and D.~Zargaryan \inst{\ref{DIAS},\ref{RAU}}
\and A.A.~Zdziarski \inst{\ref{NCAC}}
\and A.~Zech \inst{\ref{LUTH}}
\and S.J.~Zhu \inst{\ref{DESY}}
\and A.~Zmija \inst{\ref{ECAP}}
\and S.~Zouari \inst{\ref{APC}}
\and N.~\.Zywucka \inst{\ref{NWU}}
}}

\institute{
University of Namibia, Department of Physics, Private Bag 13301, Windhoek 10005, Namibia \label{UNAM} \and
Dublin Institute for Advanced Studies, 31 Fitzwilliam Place, Dublin 2, Ireland \label{DIAS} \and
Max-Planck-Institut f\"ur Kernphysik, P.O. Box 103980, D 69029 Heidelberg, Germany \label{MPIK} \and
High Energy Astrophysics Laboratory, RAU,  123 Hovsep Emin St  Yerevan 0051, Armenia \label{RAU} \and
Aix Marseille Universit\'e, CNRS/IN2P3, CPPM, Marseille, France \label{CPPM} \and
Centre for Space Research, North-West University, Potchefstroom 2520, South Africa \label{NWU} \and
Laboratoire d'Annecy de Physique des Particules, Univ. Grenoble Alpes, Univ. Savoie Mont Blanc, CNRS, LAPP, 74000 Annecy, France \label{LAPP} \and
University of Oxford, Department of Physics, Denys Wilkinson Building, Keble Road, Oxford OX1 3RH, UK \label{Oxford} \and
IRFU, CEA, Universit\'e Paris-Saclay, F-91191 Gif-sur-Yvette, France \label{IRFU} \and
Instytut Fizyki J\c{a}drowej PAN, ul. Radzikowskiego 152, 31-342 Krak{\'o}w, Poland \label{IFJPAN} \and
DESY, D-15738 Zeuthen, Germany \label{DESY} \and
Obserwatorium Astronomiczne, Uniwersytet Jagiello{\'n}ski, ul. Orla 171, 30-244 Krak{\'o}w, Poland \label{UJK} \and
School of Physics, University of the Witwatersrand, 1 Jan Smuts Avenue, Braamfontein, Johannesburg, 2050 South Africa \label{WITS} \and
Department of Physics and Electrical Engineering, Linnaeus University,  351 95 V\"axj\"o, Sweden \label{Linnaeus} \and
Institut f\"ur Astronomie und Astrophysik, Universit\"at T\"ubingen, Sand 1, D 72076 T\"ubingen, Germany \label{IAAT} \and
Laboratoire Univers et Théories, Observatoire de Paris, Université PSL, CNRS, Université de Paris, 92190 Meudon, France \label{LUTH} \and
Sorbonne Universit\'e, Universit\'e Paris Diderot, Sorbonne Paris Cit\'e, CNRS/IN2P3, Laboratoire de Physique Nucl\'eaire et de Hautes Energies, LPNHE, 4 Place Jussieu, F-75252 Paris, France \label{LPNHE} \and
Astronomical Observatory, The University of Warsaw, Al. Ujazdowskie 4, 00-478 Warsaw, Poland \label{UWarsaw} \and
Friedrich-Alexander-Universit\"at Erlangen-N\"urnberg, Erlangen Centre for Astroparticle Physics, Erwin-Rommel-Str. 1, D 91058 Erlangen, Germany \label{ECAP} \and
Universit\'e Bordeaux, CNRS/IN2P3, Centre d'\'Etudes Nucl\'eaires de Bordeaux Gradignan, 33175 Gradignan, France \label{CENBG} \and
Université de Paris, CNRS, Astroparticule et Cosmologie, F-75013 Paris, France \label{APC} \and
Department of Physics and Astronomy, The University of Leicester, University Road, Leicester, LE1 7RH, United Kingdom \label{Leicester} \and
Institut f\"ur Physik und Astronomie, Universit\"at Potsdam,  Karl-Liebknecht-Strasse 24/25, D 14476 Potsdam, Germany \label{UP} \and
School of Physical Sciences, University of Adelaide, Adelaide 5005, Australia \label{Adelaide} \and
Laboratoire Leprince-Ringuet, École Polytechnique, CNRS, Institut Polytechnique de Paris, F-91128 Palaiseau, France \label{LLR} \and
Laboratoire Univers et Particules de Montpellier, Universit\'e Montpellier, CNRS/IN2P3,  CC 72, Place Eug\`ene Bataillon, F-34095 Montpellier Cedex 5, France \label{LUPM} \and
Institut f\"ur Astro- und Teilchenphysik, Leopold-Franzens-Universit\"at Innsbruck, A-6020 Innsbruck, Austria \label{LFUI} \and
Universit\"at Hamburg, Institut f\"ur Experimentalphysik, Luruper Chaussee 149, D 22761 Hamburg, Germany \label{HH} \and
Landessternwarte, Universit\"at Heidelberg, K\"onigstuhl, D 69117 Heidelberg, Germany \label{LSW} \and
Institute of Astronomy, Faculty of Physics, Astronomy and Informatics, Nicolaus Copernicus University,  Grudziadzka 5, 87-100 Torun, Poland \label{NCUT} \and
Department of Physics, Rikkyo University, 3-34-1 Nishi-Ikebukuro, Toshima-ku, Tokyo 171-8501, Japan \label{Rikkyo} \and
Nicolaus Copernicus Astronomical Center, Polish Academy of Sciences, ul. Bartycka 18, 00-716 Warsaw, Poland \label{NCAC} \and
Institut f\"ur Physik, Humboldt-Universit\"at zu Berlin, Newtonstr. 15, D 12489 Berlin, Germany \label{HUB} \and
Department of Physics, University of the Free State,  PO Box 339, Bloemfontein 9300, South Africa \label{UFS} \and
GRAPPA, Anton Pannekoek Institute for Astronomy, University of Amsterdam,  Science Park 904, 1098 XH Amsterdam, The Netherlands \label{GRAPPA} \and
Yerevan Physics Institute, 2 Alikhanian Brothers St., 375036 Yerevan, Armenia \label{YPI}
}
\offprints{H.E.S.S.~collaboration,
\protect\\\email{contact.hess@hess-experiment.eu};
\protect\\\protect\footnotemark[1] Corresponding authors
}


  \abstract
  {Supernova remnants (SNRs) are commonly thought to be the dominant sources of { Galactic} cosmic rays up to the knee of the {cosmic-ray} spectrum at a few PeV. Imaging Atmospheric Cherenkov Telescopes have revealed young SNRs {as very-high-energy (VHE, >100 GeV)} gamma-ray sources, but for only a few SNRs the hadronic cosmic-ray origin of their gamma-ray emission {is} indisputably established. In all these cases, the gamma-ray spectra exhibit a spectral cutoff at energies 
  much below  
  {100 TeV and thus do not reach the PeVatron regime}.}
   {The aim of this work was {to achieve a firm detection} for the oxygen-rich SNR LMC N132D in the {VHE} gamma-ray domain {with an extended set of data}, and to clarify the spectral characteristics and the localization of the gamma-ray emission from this exceptionally powerful gamma-ray{-emitting SNR}.}
   {We analyzed 252 hours of High Energy Stereoscopic System (H.E.S.S.) observations towards{SNR} N132D that were accumulated between December 2004 and March 2016 during a deep survey {of} the Large Magellanic Cloud, adding {104} hours of observations to the previously published data set {to ensure a $>5\sigma$ detection}. To broaden the gamma-ray spectral coverage required for modeling the spectral energy distribution, an analysis of  \textit{Fermi}-LAT Pass 8 data was also included.} 
   {We {unambiguously detect}
   N132D at {VHE} with {a} significance of
   5.7 $\sigma$. We report the results of a detailed analysis of its spectrum and localization based on the extended H.E.S.S. data set. The joint analysis of the extended H.E.S.S and Fermi-LAT data results in a spectral energy distribution in the energy range from 1.7 GeV to 14.8 TeV, which suggests a high luminosity of N132D at GeV and TeV energies. 
   We set a lower limit on a gamma-ray cutoff energy of {8 TeV} with a confidence level of 95$\%$. The new gamma-ray spectrum as well as multiwavelength observations of N132D when compared to physical models {suggests} a hadronic origin of {the} {VHE} gamma-ray emission. }
   {{SNR N132D is a VHE gamma-ray source that shows a spectrum extending to the VHE domain without a spectral cutoff at a few TeV, unlike the younger oxygen-rich SNR Cassiopeia A.}
   The gamma-ray emission is best explained by a dominant hadronic component formed by {diffusive} shock acceleration. The gamma-ray properties of N132D may be affected by an interaction with a nearby molecular cloud that partially lies {inside} the 95$\%$ confidence region of the source position.}

\maketitle
%

\section{Introduction}

Supernova remnants (SNRs) have long been thought to be the dominant sources of Galactic cosmic rays (CRs). If this is the case, then 5--10\% of the explosion energy {needs} to be transferred to the 
CR particles to explain the CR energy budget \citep{Ginzburg:1964}. This concerns atomic nuclei (hadronic CRs) in particular, 
which make up $\sim$99$\%$ of CRs that are detected on Earth. Gamma-ray observations provide a probe of energetic particles accelerated in SNRs.
Accelerated electrons can produce gamma rays through their inverse Compton scattering and bremsstrahlung ({the} leptonic scenario),
and accelerated hadrons {can produce gamma rays} through the production of short-lived neutral pions {by} proton-proton ({or ion-ion})
interactions ({the} hadronic scenario).
In both cases, the gamma-ray photon energy is typically $\sim 10\%$ of the energy of the accelerated particle.
The highest {CR} energies to be expected in SNRs are {at least} $10^{14}-10^{15}$~eV \citep{Lagage:1983,berezhko_maximum_1996}, which should
lead to gamma-ray emission {in the TeV--sub PeV range}.

To date, {25} sources associated with SNRs\footnote{See \url{http://tevcat.uchicago.edu/}.}
have been firmly detected in the very-high-energy gamma-ray regime \citep[VHE, 100 GeV $<$E$<$ 100 TeV, ][]{HESSSNR_population_2018}, some of which were even
first discovered in {VHE} gamma rays \citep[e.g.,][]{HESSSNR_2018}.
These {VHE} gamma-ray detections were {achieved with} {ground-based Imaging Atmospheric Cherenkov Telescope (IACT) arrays.}
Thirteen of these sources have been identified as Galactic shell-type SNRs, ten additional sources are SNRs interacting with molecular clouds, and a few remaining sources are composite SNRs, that is, they also host a pulsar wind nebula (PWN).

In the high-energy regime (HE, {100 MeV} $\lesssim$E$\lesssim$ 100 GeV), $\sim$30 gamma-ray sources are classified as SNRs in the
\textit{Fermi}-LAT SNR catalog \citep{acero_first_2016}. These include the mature SNRs {($\gtrsim$5000~yr)} IC443 and W44, whose gamma-ray spectra display a clear
pion bump signature, which provides evidence for a hadronic origin of the gamma-ray emission \citep{Ackerman:2013}. {The outer shocks of} these two SNRs are interacting with dense molecular clouds,
which act as reservoirs of target material for pion production.
{Their} gamma-ray spectra show spectral breaks, suggesting that the highest-energy {accelerated} protons have largely escaped the SNR {shell}. 
Further evidence {that {VHE} emission in mature SNRs originates from CRs that have
escaped the SNR} is provided by
{VHE} gamma-ray observations of W28, {for which} {VHE} emission
was detected from {a} nearby molecular cloud situated outside the shell of the
SNR \citep[e.g.,][]{HESS_W28paper, Gabici2017}.

In general, gamma-ray studies show that
(i) several young SNRs ($\lesssim$1000~yr) have prominent {VHE} gamma-ray emission,
but there is no evidence that the underlying CR spectrum extends to the CR knee at $3\times 10^{15}$~eV;  the bright young SNR, Cassiopeia A ({Cas A,} $\sim$340~yr), exhibits a cutoff at a surprisingly low energy of 3.5 TeV 
in the gamma-ray spectrum \citep{CasA_magic:2017};
(ii) mature SNRs evolving in dense regions are bright GeV sources, but exhibit a cutoff and/or break in their HE spectra, and they are therefore not  prominent {VHE} gamma-ray sources;
(iii) for some mature SNRs, it is suggested that the gamma-ray emission is caused by CRs that have escaped the
 SNR shell and are now colliding with dense nearby gas.
 
{{The} SNR N132D {is a remarkable object} located in the Large Magellanic Cloud (LMC).  
At a distance of $\sim$50 kpc \citep[e.g.,][]{Pietrzynski2019}, N132D is the only known extragalactic {VHE} emitting SNR \citep{collaboration_exceptionally_2015}.
{SNR} N132D was discovered in gamma rays in the HE and {the} {VHE} regimes \citep{collaboration_exceptionally_2015, FermiLAT_LMC_2016}. 
{Its} gamma-ray luminosity was {estimated to be about}  10$^{36}$ erg s$^{-1}$ {, making it}{ one of the}
most luminous {VHE} gamma-ray SNRs {detected so far.} {It is the brightest X-ray SNR in the LMC of 38 cataloged LMC SNRs} \citep{Maggi:2016}.
{However, unlike many other {VHE}-emitting {shell-type} SNRs,   N132D shows only strong thermal X-ray emission} \citep[e.g.,][]{mathewson_supernova_1983, hughes_asca_1998,behar_high-resolution_2001,borkowski_x-ray-emitting_2007} without}
 evidence {for an X-ray synchrotron component}
\citep[e.g.,][]{bamba_transition_2018}. It is also a luminous source in the radio \citep{dickel_radio_1995} and infrared (IR) bands \citep{seok_survey_2013}.
There is evidence that N132D was caused by a powerful explosion {with an energy} of (3-5)$\times 10^{51}$ erg \citep{dickel_radio_1995,hughes_asca_1998},
three to five times more {than the} canonical supernova explosion energy.

N132D has a peculiar horseshoe morphology {in the X-ray band {\citep[see, e.g.,  Fig. 2 in ][]{borkowski_x-ray-emitting_2007}} and also in the radio and IR bands}, with a typical {angular} size of $1.4^{\prime}$ by $1.8^{\prime}$. {This corresponds to a physical size of about} 20~pc by 26~pc.
{The} northeast region of the SNR is fainter {than the southwest region,
which suggests that the SNR interacts with  denser material in the southwest.}
Molecular clouds are projected toward the southwestern region {of the remnant} \citep[e.g.,][]{braiding_mopra_2018}. 
{HI emission also shows that N132D is surrounded by relatively
dense HI gas, with a gradient toward the southwest. \citet{Sano:N132D} estimated
an HI density as high as $n\approx 30$~cm$^{-3}$.
}
{It seems  plausible that} in the southern part of the shell,  molecular cloud material has been swept up and heated by a high-speed shock, resulting in the peculiar morphology of N132D \citep[see][]{williams_dust_2006,tappe_shock_2007,tappe_polycyclic_2012}.
{\citet{Sano:N132D} also reported that there are small clouds
in the south and in the center of N132D that shows signs of {present or past interaction} with the shock.}

The age of N132D is estimated to be $\sim$2500 years \citep{Vogt_Dopita_2011}. {It is} therefore one of the oldest {VHE} gamma-ray-emitting shell-type SNRs.
It is much older than Cas A ($\sim$340~yr), to which it is often 
compared in other aspects: both SNRs {probably have evolved in a} wind-blown bubble, and both contain oxygen-rich ejecta,
which suggest that their 
explosions were the result of  {core-collapse supernovae of} massive stars ($\gtrsim 18$~M$_\odot$, at the zero-age main sequence).
The age of another oxygen-rich SNR, Puppis A, is similar to that of N132D, but it has not yet been  detected in {VHE} gamma rays although it is located only 1.3 kpc away \citep{PuppisA:2015}. 
The difference in {VHE} gamma-ray properties of
N132D, Cas A, and Puppis A makes a detailed study of N132D relevant for the important
question of the timescale on which the bulk of CRs are accelerated, and of the timescale on which the CRs
eventually escape the shells as a function of CR energy.

\cite{collaboration_exceptionally_2015} reported a gamma-ray excess from N132D at a statistical significance level of 4.7 $\sigma$ and a spectrum extending up to 10 TeV, based on H.E.S.S. observations with an exposure time of {148} hours.
{Our goal is to} update this result by reporting the analysis of H.E.S.S. observations with 252 hours of total exposure time. The HE gamma-ray counterpart of {VHE} emission {is also studied} with \textit{Fermi}-LAT Pass 8 data. 
{This joint study in the HE and VHE gamma-ray bands allows us to place new constraints on the mechanism}
of gamma-ray production and the {underlying} 
population{s} of
accelerated particles responsible for the gamma-ray emission.

The content of this paper is {structured} as follows:
In  section \ref{sec:obs} the data analysis of both H.E.S.S. and \textit{Fermi}-LAT observations {are described}.
In Section \ref{sec:results} the H.E.S.S {VHE} spectrum
and gamma-ray morphology {are presented}.
In Section \ref{sec:SED} a multiwavelength spectrum modeling {is explained}. 
These results are discussed in Section \ref{sec:discuss} in the context of the environment of N132D as
well as in the context of the {VHE-emitting} SNR population. 
The conclusions of this study are presented in  
Section \ref{sec:conclusion}.

\section{Observations and data analysis}\label{sec:obs}

\subsection{H.E.S.S. experiment}
The High Energy Stereoscopic System\footnote{ https://www.mpi-hd.mpg.de/hfm/HESS/} (H.E.S.S.) is an array of five {IACTs} located in the Khomas Highland {in} Namibia at an altitude
of 1800 m above sea level. It has been operating since December 2003. In its first phase ("H.E.S.S. I"), the array consisted of four identical
12m Cherenkov telescopes (CT1-4) placed at the corners of a square of 120m {side length}. CT1-4 are equipped with mirrors with a total area of
107 m$^2$ and cameras {with} 960 photo-multiplier tubes each. The stereoscopic IACT technique allows determining the energy and direction of VHE gamma rays through imprints of the Cherenkov light emitted by secondary charged particles of {a} shower
initiated by a primary VHE photon entering the atmosphere.
With {the 5$^{\circ}$ field of view (FoV) of} CT1-4, gamma rays can be reconstructed with a typical angular
resolution of $\sim$0\fdg1 {(68$\%$ containment radius)}. {The} energy threshold {is} $\sim$100 GeV for small zenith angles and below 1 TeV for zenith angles up to 60$^{\circ}$, and the energy resolution is  $\sim$15$\%$ \citep[for further details, see][]{Aharonian:2006}. {The} fifth Cherenkov telescope (CT5) with
a diameter of 28 m ({mirror} area {of} 614 m$^2$) was added in the center of the {CT1-4} array in July 2012, initiating the second phase of the experiment, "H.E.S.S. II". 
This study mostly uses {H.E.S.S. I} observations because{ only about} $5\%$ of the {data} were taken by the full CT1-5 array.
H.E.S.S. performance {depends} on the zenith angle, configuration cuts, {and} the number of telescopes considered in the reconstruction {(for more details, see 
sections \ref{sec:data-set} and \ref{sec:analysis}).}

\subsection{Data set: LMC survey}
\label{sec:data-set}

Its {location} in the southern hemisphere makes H.E.S.S. the only currently operating IACT array able to observe the LMC in the {VHE} band. 
With its angular size of $\sim$8$^{\circ}$, the
LMC galaxy is compact enough for the H.E.S.S. telescopes to perform a survey.
Such a survey was conducted over 13 years, from 2004 {to} 2017, during which the full LMC galaxy was covered with over 325 hours of observations. {The exposure of the LMC is rather
inhomogeneous and} mostly concentrated on the central region around the Tarantula nebula.
H.E.S.S. observations from 2003 to 2012 led to the discovery of three individual VHE gamma-ray-emitting sources,
including SNR N132D, {along with the PWN of N 157B and the superbubble 30 Dor C} \citep{HESSN157B_2012,collaboration_exceptionally_2015}, as well as the detection of the gamma-ray binary LMC P3 \citep{HESS_LMCP3}.
The data set used for our analysis 
was taken between December 2004 and March 2016. It consists of 252 hours of total exposure time, adding {104} hours to the previously published data set.
The zenith angle of these observations {spans a range of 45-57$^{\circ}$, with a mean of 46$^{\circ}$,}
and a mean azimuth angle of 182\fdg5. The data set was selected with a maximum offset of 2\fdg5 around 
the source to avoid observations with large offset angles
with respect to the camera center. {For N132D, the observation positions were mostly toward the east side of the SNR.} 
{This is due to the nature of the observations, which are part of a survey, instead
of being dedicated to N132D: the pointings were mostly directed toward N157B in the Tarantula} {nebula}, which 
is situated toward the east of N132D. This results in offset angles {in the range} of 0\fdg16 to 2\fdg5 and a mean offset angle 
for {N132D} of 1\fdg13.
{The observations} are thus taken at rather large offset, for which
the acceptance is reduced compared to sources located closer to the camera center.  
Observations at large zenith angles with IACTs in general result in  a lower effective area
at energies below 1 TeV, but an increase in effective area
above 1 TeV \citep[e.g.,][]{Magic_Crab}. 
 {Therefore}  these observations were particularly
well suited for measuring the high-energy {part} of the N132D gamma-ray spectrum.

\subsection{H.E.S.S. data analysis}
\label{sec:analysis}

{Data corresponding to 252 hours} were recorded in 28-minute exposures that are {called} runs. Each calibrated run that passed the quality criteria of a dead time shorter than 30$\%$ of the run duration, a low fluctuation in trigger rate, and {acceptable} weather conditions  \citep[see][]{Aharonian:2006,hahn2014impact} was analyzed using the ImPACT
framework described {by} \cite{ImPACT_paper}, which includes a boosted decision-tree-based event classification algorithm to distinguish
gamma rays from the charged particle background \citep{Ohm_2009}.
{This analysis was performed using a special configuration that} 
selected events with a minimum of three {shower images} {in} different telescopes.
Using this {event} selection cut {allowed us} {to improve} the signal-to-background ratio for this faint gamma-ray source {in the presence of} a strong diffuse component, or multiple unresolved sources constituting an astrophysical background, which {may be} present in the vicinity of N132D. 

\begin{figure}[ht]
    \centering
    \includegraphics[width=0.49\textwidth]{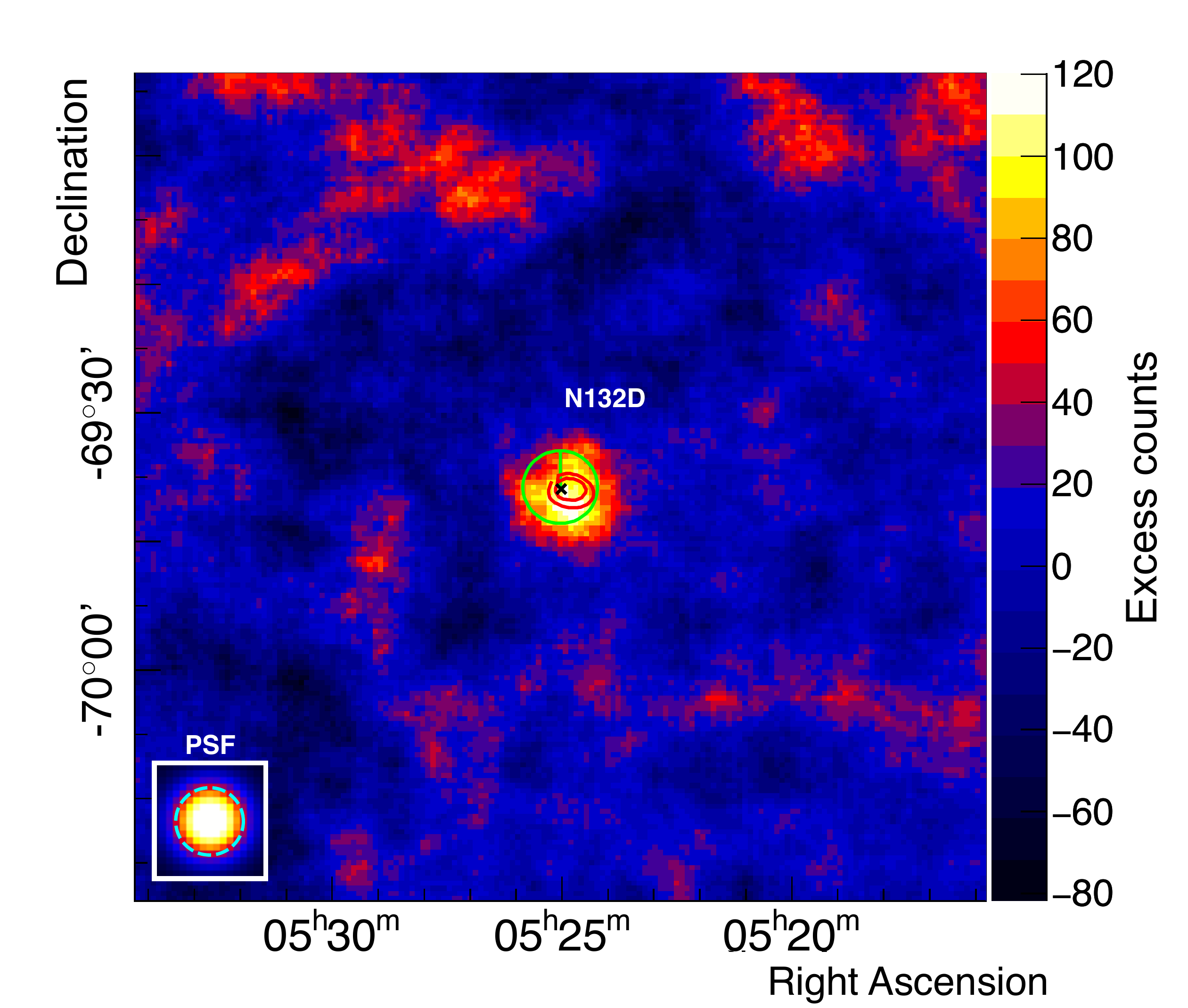} 
    \caption{H.E.S.S. excess map with a size of 1\fdg6, smoothed with a {Gaussian} of width 0\fdg1. The ON region with a radius of 0\fdg07, centered on {the} N132D X-ray source {position,} is shown in green, and the {2 and 3{$\sigma$} of the {best-fit} position (see section \ref{sec:morpho}) are represented in red.}} 
    \label{fig:Excess_map}
\end{figure}

To compute the maps and statistics presented in section \ref{sec:results}, ON events were selected from
a circle centered {on the geometrical center of the  X-ray emission (referred to as the ``X-ray source position'') at RA  $5^\mathrm{h}25^\mathrm{m} 2.2^\mathrm{s}$, Dec $-69^{\circ}38^{\prime}39^{\prime\prime}$. OFF events were selected according to a {ring-background} method \citep[][]{Berge:2007}.}
An additional cut on the energy of reconstructed gamma-{rays} was made above 1.3 TeV 
in order to {improve the signal-to-background ratio for this hard gamma-ray source and} keep only gamma-ray events from a well-{determined} direction because the angular 
resolution improves by 10$\%$  from {800 GeV} to 1.3 TeV using the ImPACT shower reconstruction software. This resulted in a point spread function (PSF) of 0\fdg06, {which corresponds to the 68$\%$ containment radius of the fit on the observation position}. The event count excess was computed using N$_{excess}$=N$_{on}$ - $\alpha$N$_{off}$, with $\alpha$ the normalization factor for the OFF area with respect to the ON area. 
The cumulative statistical significance was {calculated} using Equation\,17 of \cite{Li:1983}. {The excess map is shown in Fig.~\ref{fig:Excess_map}.}

For the spectrum, the {reflected-background} method was applied with {a} three-telescope-image selection cut and {a low-energy} cut above 1.3 TeV.
The spectral background was derived from background control regions that were defined run-wise and were chosen to have the same offset
to the camera center as the source region to ensure a nearly identical spectral response. For maps and {the} spectrum, {the} sky areas with known sources or with excess events above a certain significance threshold were excluded. 
For reliability, the results {were} cross-checked {with} an independent analysis and calibration chain
\citep{DeNaurois:2009}.

\subsection{\textit{Fermi}-LAT data analysis}
\label{sec:Fermi_analysis}

The Large Area Telescope (\textit{Fermi}-LAT) is a high-energy gamma-ray
telescope installed on the Fermi spacecraft \citep[][]{atwood09}. 
It uses a pair-conversion technique for gamma-ray photon detection, and has a large FoV {of} about 20\% of the sky. 
The \textit{Fermi}-LAT has been scanning the entire sky since August 2008, recording astrophysical gamma rays with energies in the range 30 MeV to 300 GeV. 
Despite the uniform coverage of the entire LMC galaxy and the excellent
spectral capabilities of \textit{Fermi}-LAT, the previous analyses
of \textit{Fermi}-LAT observations {were inconclusive} regarding the origin of {gamma-ray} emission from N132D. {In particular,} 
different values of {the photon spectral} index were reported{:} 
$\Gamma=1.4\pm0.3$ {in} \citet[][]{FermiLAT_LMC_2016} and
$\Gamma=2.07\pm0.19$ in the LAT 8-year source catalog \citep[the
4FGL catalog;][]{Very_new_4FGLcat}. In the latter work, the source 4FGL J0524.8-6938 was identified with N132D, and it was argued that the derived photon index of 2.07 supports the hypothesis of dominantly hadronic emission, while for SNRs exhibiting gamma-ray spectra with
very hard photon indices, {$\Gamma\leqslant 1.6$,} the gamma-ray emission is most likely of leptonic origin \citep[][]{Very_new_4FGLcat}.

\begin{figure}[ht]
\centering
\includegraphics[width=0.49\textwidth, angle=0]{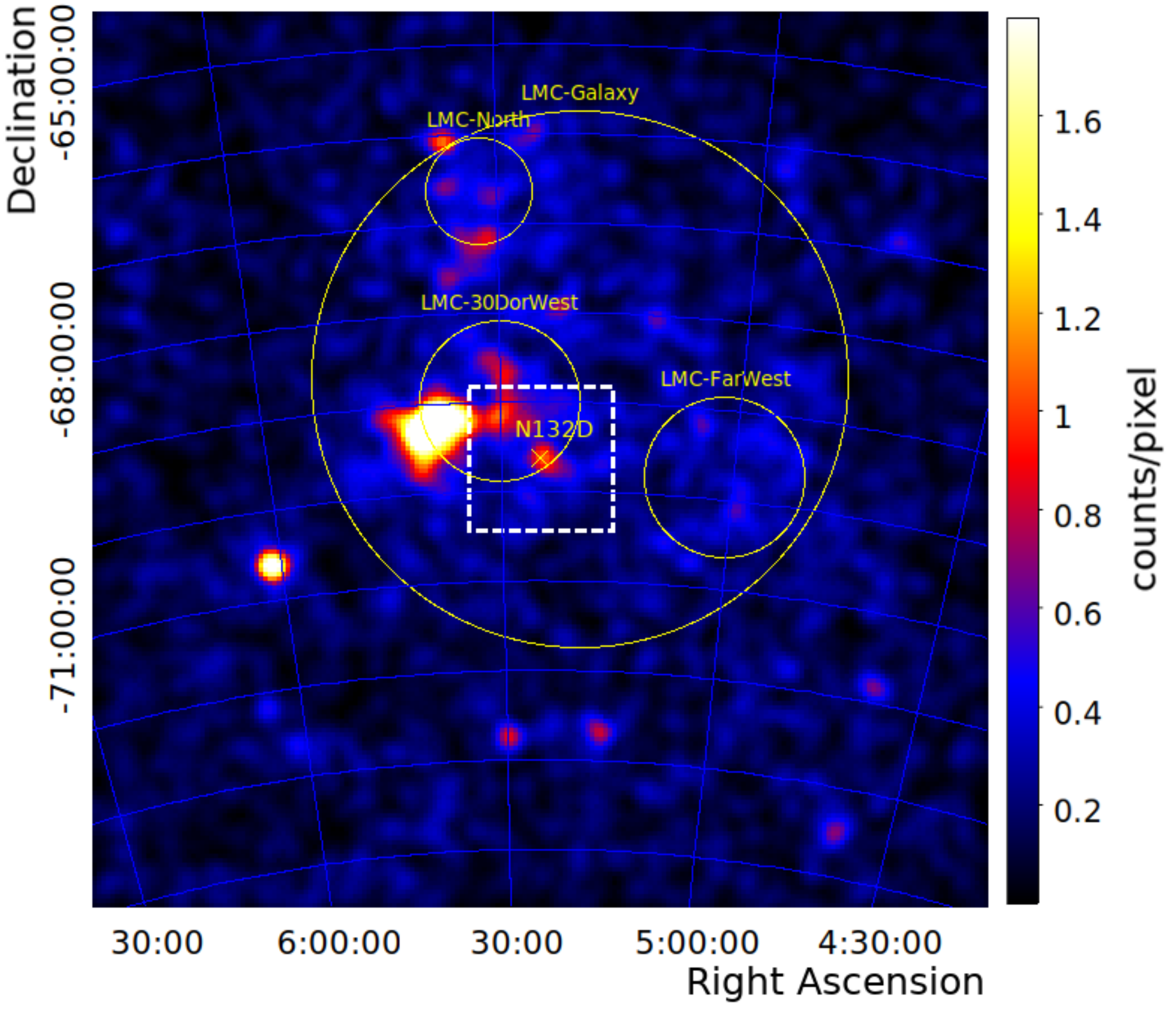}
\caption{Gaussian ($\sigma$=0\fdg2) kernel smoothed \textit{Fermi}-LAT
count map centered on the position of N132D for the energy range 3-250 GeV.
The central dashed box corresponds to the {FoV} of the H.E.S.S. excess map shown in Fig. \ref{fig:Excess_map}. {The bright \textit{Fermi}-LAT sources at the eastern edge of the LMC-30 DorWest region are N157B and PSR B0540-69. Their emissions overlap at this resolution.}}
\label{fig:fermicmap}
\end{figure}

Pass 8 R3 \texttt{SOURCE} class photon data\footnote{https://fermi.gsfc.nasa.gov/ssc/data/} (\texttt{evclass}=128)
spanning 10.8 years between August 4, 2008, and May 15, 2019,
with energies between 1 GeV and 250 GeV were selected. For the data
analysis, the \texttt{FERMITOOLS} v1.0.7 package and
\texttt{P8R3\_SOURCE\_V2} instrument response functions were used.
Contamination from the gamma-ray-bright Earth limb was avoided by
removing all events with zenith angle larger than 90$^{\circ}$. The
recommended quality cuts \texttt{(DATA\_QUAL>0 \&\& LAT\_CONFIG==1)}
were applied. The $10^{\circ}\times10^{\circ}$ square region of interest
with spatial bins 0\fdg05 in size was {centered} on the position of
N132D. A \textit{Fermi}-LAT count map in the range 3-250 GeV
smoothed with a Gaussian kernel is shown in Fig.~\ref{fig:fermicmap}. The choice of the low-energy limit at 3 GeV was made for illustration purposes because it {provides} a narrower point spread function.
{The central region of 1\fdg6$\times$1\fdg6 in Fig.~\ref{fig:fermicmap} indicated with the dashed line corresponds to the H.E.S.S. excess map shown in Fig.~\ref{fig:Excess_map}.}
Given the diameter of {$\sim$1.6$^{\prime}$} of N132D in X-rays, a point-like source model for its counterpart in HE gamma rays {was adopted}.
Gamma-ray sources from the 4FGL catalog {within a 17$^{\circ}$ radius from N132D} were included to model the data. To model the diffuse emission from the LMC, 
the spatial templates {for four diffuse sources,} LMC-FarWest, LMC-Galaxy, LMC-30DorWest, and LMC-North, 
provided by the \textit{Fermi}-LAT collaboration \citep[][]{FermiLAT_LMC_2016}, were used. These sources 
are outlined {with} circles in Fig.~\ref{fig:fermicmap}. To model the Galactic and isotropic background diffuse emission, the standard templates \texttt{gll\_iem\_v07.fits} and
\texttt{iso\_P8R3\_SOURCE\_V2\_v1.txt} were used. The spectral shapes
of the sources were taken from the 4FGL catalog. The normalization and the photon index of N132D were derived from the likelihood
analysis along with the normalizations of four strong point sources, 
including 4FGL J0537.8-6909 and 4FGL J0540.3-6920, and all diffuse sources, 
while the normalizations of fainter point sources were held fixed at 
the 4FGL catalog values. 
{A} binned likelihood analysis was applied {to} the data 
using the Fermi Science tool \texttt{gtlike}. 
{To assess the goodness of fit, test-statistic and residual maps were produced showing some residuals at the position of {SNR} N63A. However, the inclusion of a new source corresponding to N63A in the model negligibly affects the derived spectrum of N132D.} 
{To extract a spectral energy distribution (SED),} the photon index of N132D {derived
to be $1.86\pm0.25$ was fixed at the central value}, the data were rebinned in {four} broad logarithmically 
spaced bands in energy, and a binned likelihood analysis was applied {to} the data 
in each of these bands.

The main analysis with {all} LMC templates {was first performed, providing} the best estimate of {the} SNR N132D HE emission. A second analysis {was then launched} to estimate the uncertainties related to the presence of a molecular cloud near N132D. {For this purpose,} LMC-Galaxy and LMC-30DorWest {templates} were modified by patching
the area enclosing the molecular cloud, with {constant} intensities matching 
those in the nearby region and {keeping their initial spectral shapes}. Given the proximity of N132D and the molecular cloud, the alternative 
model {predicts} a stronger HE gamma-ray signal from N132D and leads to an expected VHE flux that is higher by about 30\%, which is an estimate of the {HE emission of SNR N132D plus the nearby molecular cloud}. The data points 
obtained from the \textit{Fermi}-LAT analysis are shown in Fig. \ref{fig:HESS_Flux} for the main model.

\section{Results}\label{sec:results}
\subsection{Statistics and spectral properties}
The H.E.S.S analysis of N132D reveals an excess of 95 gamma-ray events {above an energy of 1.3 TeV}, corresponding to a significance of 5.7 $\sigma$ for an exposure of 252 hours. 
The {VHE} gamma-ray emission from N132D is thus unambiguously detected{, for which there was previous evidence from H.E.S.S.
at a lower {significance} level of 4.7 $\sigma$ \citep{collaboration_exceptionally_2015}.}

The excess corresponds to 337 ON events and 11179 OFF events, and a normalization parameter $\alpha$= 0.0216 established with the  
{ring-background} method with standard settings. ON events were extracted in a circle of radius 0\fdg07 
centered on the X-ray source position. 
The corresponding excess map is shown in Fig. \ref{fig:Excess_map}.

\begin{figure*}[ht]%
    \centering
     {\includegraphics[width=0.47\textwidth]{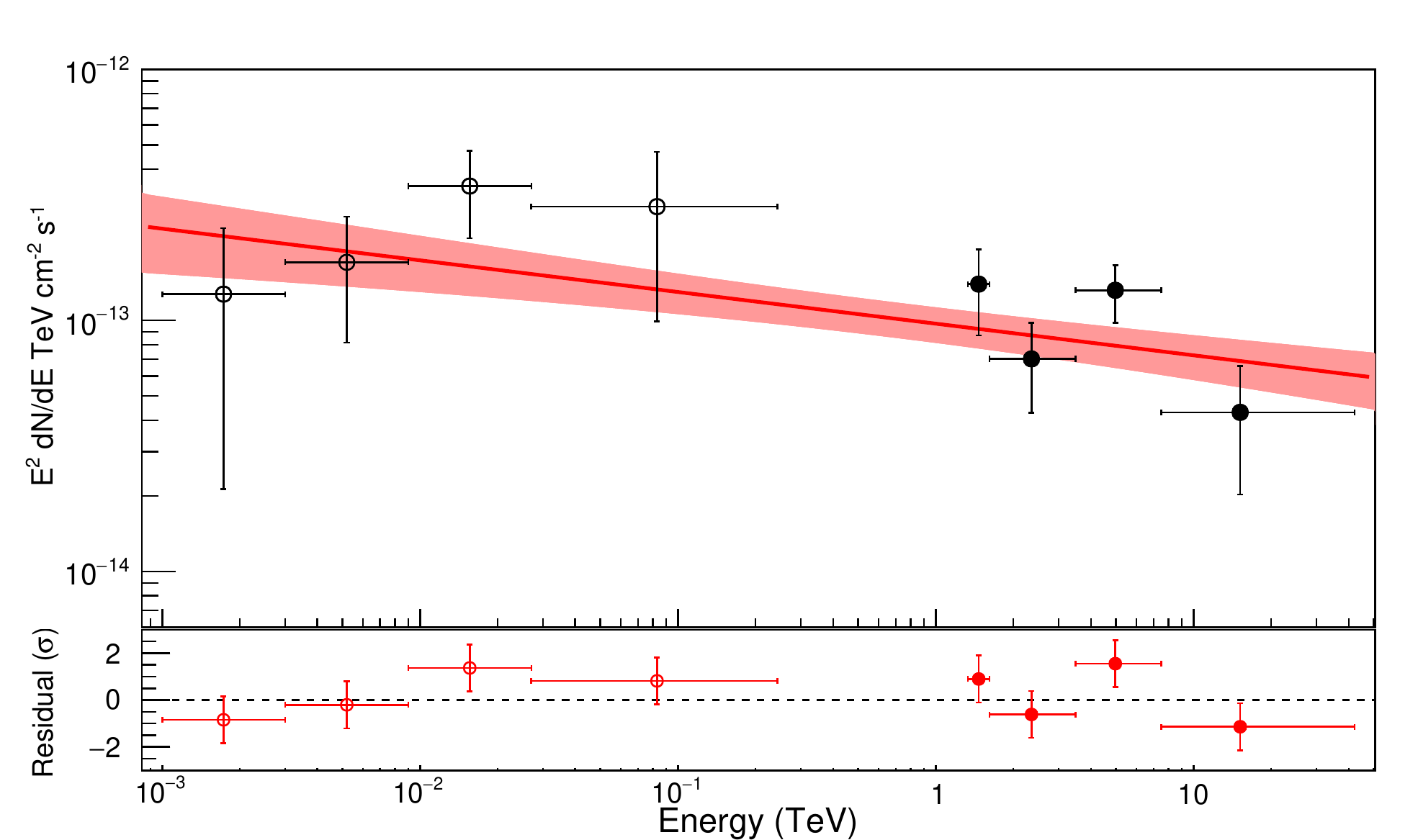}}
     \qquad
     {\includegraphics[width=0.47\textwidth]{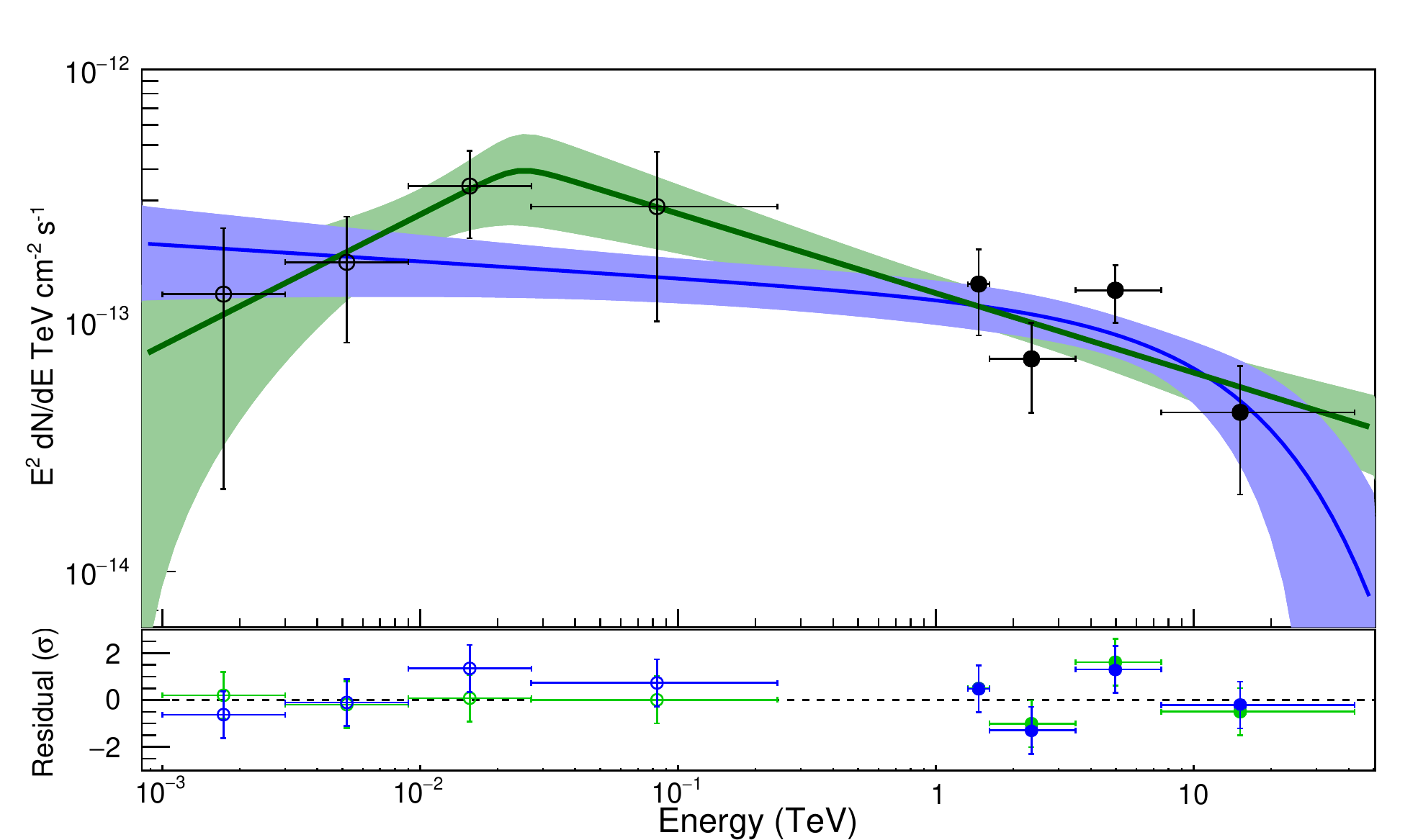}}
    \caption{Left: \textit{Fermi}-LAT (open circles) and H.E.S.S (dots) combined spectrum fit with a power law. Right: \textit{Fermi}-LAT and H.E.S.S combined spectrum fit with a power law with an exponential cutoff (blue) and a broken power law (green). 
    For both figures, the {best-fit} functions are shown with 68 $\%$ error bars in the shaded region, as well as residuals compared to the fit. {The flux point error bars represent the statistical uncertainties.} The fit parameters are reported in Table \ref{tab:stats}. }
    \label{fig:HESS_Flux}
\end{figure*}

Fig.~\ref{fig:HESS_Flux} shows the combined H.E.S.S. and \textit{Fermi}-LAT photon energy spectrum spanning an energy range of 1 GeV to 40 TeV, 
where H.E.S.S. data points {were chosen to} start at 1.3 TeV.  
{The H.E.S.S.} spectrum 
was extracted using the same analysis cuts as for the maps and the {reflected-background}
method. Spectral points between 1.3 and 40 TeV were binned in energy bands {requiring} a minimum of 2 $\sigma$ significance
per point. 
{The} H.E.S.S. spectrum exhibits a 2 $\sigma$ significant spectral bin at $\sim$ 15~TeV.

The H.E.S.S. spectrum is well fit with a single power law ($dN/dE = \Phi_0\times (E/1\mathrm{TeV})^{-\Gamma}$) with an index
$\Gamma = 2.3\pm 0.2$ and a differential flux normalization at 1 TeV of
$\Phi_0 = (1.31\pm0.43)\times 10^{-13} ~\mathrm{TeV^{-1}cm^{-2}s^{-1}}${, with} a $\chi^{2}$ of 3.9 for 2 degrees of freedom. The derived photon index is consistent with the previous estimate by H.E.S.S., yielding $\Gamma = 2.4\pm 0.3$ \citep{collaboration_exceptionally_2015}. {Systematic errors on the spectrum parameters are estimated to be $\pm$ 0.3 for $\Gamma$, and $\pm$ 30$\%$ for $\Phi_0$.}{ These systematic uncertainties arise because the data set spans 13 years and the LMC is mostly observed during the rainy season, leading to {significant variations in atmospheric conditions}} \citep[see][]{N157B_2012}.
The corresponding luminosity, assuming a distance $d=50$ kpc to the source, is
$L(1 -10~\mathrm{TeV}) = (1.05 \pm 0.29) \times 10^{35} (d\mathrm{/50kpc)^{2}~erg~ s^{-1}}$.
This value is consistent with previous results and is comparable to the luminosity of the brightest {SNRs} {in the VHE band}: N132D is as luminous as HESS J1640-465 \citep{J1640_erratum2014}
and $\sim$28 times more luminous than Cas A \citep{CasA_discovery}.

{A combination of the H.E.S.S and \textit{Fermi}-LAT data allowed us to investigate the spectral properties in a wide energy range.}
The fits were performed on H.E.S.S and Fermi-LAT eight spectrum points, as shown in Fig.\ref{fig:HESS_Flux}.
The fit parameters are reported in Table \ref{tab:stats}. The combined spectrum can be perfectly fit with a {single} power law, resulting in a reduced $\chi^{2}$ of 8.2 for 6 degrees of freedom (Fig. \ref{fig:HESS_Flux} left). 
{The derived photon index,}
$\Gamma = 2.13\pm 0.05$, is consistent with the {index obtained with the} H.E.S.S. {data} alone. 
The fit {model} with a power law and an exponential cutoff (Fig. \ref{fig:HESS_Flux} right), 
$dN/dE = \Phi_0\times (E/1\mathrm{TeV})^{-\Gamma} \exp(-{E/E_{\mathrm{c}}})$, is {not preferred over} a fit with a simple power law
as derived from this table by means of a {likelihood ratio test} ($1.4\sigma$, $\Delta\chi^{2}=1.8$ with 1 degree of freedom). {Considering the Akaike information criterion \citep[AIC;][]{akaike74}, an AIC value of 12.4 is obtained while fitting a power law with an exponential cutoff to the data, which is 12.2 when a power law is fit to the data. The relative likelihood defined as $\exp((\mathrm{AIC}_{\mathrm{min}}-\mathrm{AIC}_{\mathrm{max}})/2)$ shows that the power-law model is sufficient to describe the data.}
The resulting cutoff energy is quite high, exhibiting large errors ($E_{c}$ =19$^{+60}_{-10}$ TeV). A cutoff energy lower than 8 TeV can be excluded at a 95$\%$ {confidence level (CL)}. {The energy cutoff value is based on the statistical analysis and does not take into account the H.E.S.S. systematic error in the energy reconstruction of $\sim15\%$.}
A broken power-law fit to the joint Fermi/H.E.S.S. data yields values (Fig. \ref{fig:HESS_Flux} right) of the break energy between 8 GeV and 140 GeV, with a best-fit value of 24 GeV, {and it is as likely as} a simple power law fit {($1.5\sigma$, $\Delta\chi^2=4.0$ with 2 degrees of freedom, {and an AIC value of 12.2)}}.
These lower and upper bounds were established by fitting the data with a fixed value for the energy break until a $\Delta \chi^2=1$ was reached (68$\%$ CL), leaving all other parameters free.

\begin{table*}[ht]
\centering
\label{tab:stats}
\begin{tabular}[t]{lccccc}
\midrule
 Data  &$\chi^{2}$/ndof&\multicolumn{2}{c}{$\Gamma$}      &   $\Phi_0$ (1 TeV)    & $E_{c}/E_{b}$\\
\multicolumn{4}{c}{}&$\times10^{-14} \mathrm{TeV^{-1} cm^{-2} s^{-1}}$& \\
 \midrule
 H.E.S.S. PL &3.9/2&\multicolumn{2}{c}{2.32 $\pm$ 0.22}&13.1 $\pm$ 4.3&--\\
\textit{Fermi}-LAT + H.E.S.S PL&     8.2/6 &   \multicolumn{2}{c}{2.13 $\pm$ 0.05} &   9.7 $\pm$ 1.6  &   --   \\                  \textit{Fermi}-LAT + H.E.S.S ECPL        &6.4/5 &  \multicolumn{2}{c}{2.08 $\pm$ 0.06}     &   12.3 $\pm$ 2.9  &   $E_{c}$ = 19$^{+60}_{-10}$     TeV\\
\textit{Fermi}-LAT + H.E.S.S BPL    &   4.2/4 &  1.47 $\pm$ 0.43  &   2.31 $\pm$ 0.14 & 12.8 $\pm$ 2.9   &$E_{b} $= 24$^{+116}_{-16}$ GeV   \\
\bottomrule
\end{tabular}
\caption{Spectral fit parameters for the H.E.S.S. and \textit{Fermi}-LAT plus H.E.S.S. data {sets}. PL stands for "power law", ECPL for "power law with an exponential cutoff" , and "BPL" for "broken power law" (see text for details).}
\end{table*}

\subsection{Morphology in the VHE gamma-ray band}
\label{sec:morpho}
The position of the {VHE} gamma-ray source was obtained from a two-dimensional fit of the H.E.S.S. so-called {ON} map 
{consisting} of {gamma-ray-like} events in the source {region} and a background map, using the gamma-ray events selected in the OFF region 
multiplied by a scaling factor map in order to reach the source exposure level. The fit was performed using the \textit{Sherpa} package \citep{sherpa:2001}, with a log-likelihood approach \citep[{Cash} statistic;][]{Cash:1979} and the Levenberg-Marquardt algorithm \citep{levmar:1978}.
{To obtain the source position, a Gaussian source was convolved with the H.E.S.S. analysis PSF
(0\fdg06)  multiplied by the exposure map to take any asymmetry in the exposure across the analyzed sky region into account, and with the background amplitude fixed to 1. As there is no evidence for an extension, a point source assumption was used by fitting the two-dimensional Gaussian profile with a width of
 $\sigma=0\fdg005$.}
The fit results in a {best-fit} position of {RA  $5^\mathrm{h}24^\mathrm{m} 47^\mathrm{s} \pm 6.9^\mathrm{s}_{\mathrm{stat}}$ and Dec $-69^{\circ}38^{\prime}50^{\prime\prime}\pm 29^{\prime\prime}_\mathrm{stat}$} at about $1^{\prime}$ from the X-ray source position. The uncertainty of the telescope pointing leads to a systematic error of $\sim$20$^{\prime\prime}$ {per axis}. The systematic error arising from the asymmetrical exposure is negligible. The best-fit position and the confidence region {are} shown in Fig.~\ref{fig:Excess_map} and are consistent with the
center of N132D as seen in X-rays within $\sim$1.2 $\sigma$.

{Given the diameter of N132D in X-rays of $1.6^{\prime}$ and the presence of a nearby molecular cloud at a distance of  {$\sim2$} arcminutes, it is likely that the gamma-ray source has a structure on the arcminute scale. With the reported data, a two-dimensional fit with a Gaussian profile centered on the best-fit position provides only a {99}$\%$ CL upper limit on the source extension in gamma rays  corresponding to} $\sigma_{\mathrm{2D, Gauss}}=3.3^{\prime}${ (49 pc). Because the {PSF} is $3.6^{\prime}$, the localization of the emitting region is inside this radius.}

\section{Modeling the multiwavelength spectrum}\label{sec:SED}

{The results were analyzed} in the context of the multiwavelength studies of N132D.  
Radio observations at wavelengths 3 cm and 6 cm by \cite{dickel_radio_1995} {with the Australia Telescope Compact Array (ATCA)}, and nonthermal X-ray upper limits in the 0.2 to 4 keV band \citep{hughes_asca_1998} and in the 2-10 keV band provided by \cite{bamba_transition_2018} {were added to the gamma-ray spectrum.}
This led to a multiwavelength spectrum spanning {15} {orders} of magnitude (see Fig. \ref{fig:SED_leptonic}).
Radio and {nonthermal X-ray emission} can be related to gamma-ray emission {based on {two}} different scenarios. The leptonic scenario assumes that {a single population} 
of CR electrons {generates} synchrotron emission in the radio to X-ray range and inverse Compton {emission in} {gamma rays}. Alternatively, {in the hadronic scenario,} CR electrons are only responsible for the {synchrotron} radio and X-ray emissions, 
while CR hadrons {produce} {gamma-ray} emission through neutral {pions} that decay into two photons.
In this case, a ratio of $n_{\rm e}/n_{\rm p}$ {for CRs} is one of the parameters, for which the canonical value is 1/100, based on the ratio 
of{} electron-to-proton CRs measured near the Earth \citep[see, e.g.,][]{Longair:2011,KatzWaxman:2008}. {However, 
the  $n_{\rm e}/n_{\rm p}$ ratio {at} Earth may not reflect the ratio in 
SNRs, and there may be  variations in SNRs.}
Here these radiation processes {were modeled}
using {theoretical} frameworks developed {by} \cite{aharonian_angular_2010}
for the synchrotron radiation, {by} \cite{khangulyan_simple_2014} for the inverse Compton mechanism, by \cite{baring_radio_1999} for
bremsstrahlung emission, and by \cite{kafexhiu_parametrization_2014} for the pion decay.
{To model the multiwavelength spectrum of N132D {for the two scenarios,}
the Naima package was employed \citep{naima}.}
{ We used a single-zone model. In reality, emission may arise from a number of regions, such as the shocked ejecta, the main shell of the shocked ambient medium, or even, as we discuss below, from cosmic rays interacting with nearby molecular clouds. }

\subsection{Purely leptonic model}

\begin{figure*}[ht]
\centering
  \begin{tabular}{@{}cc@{}}
    \includegraphics[angle=0, width=.47\textwidth]{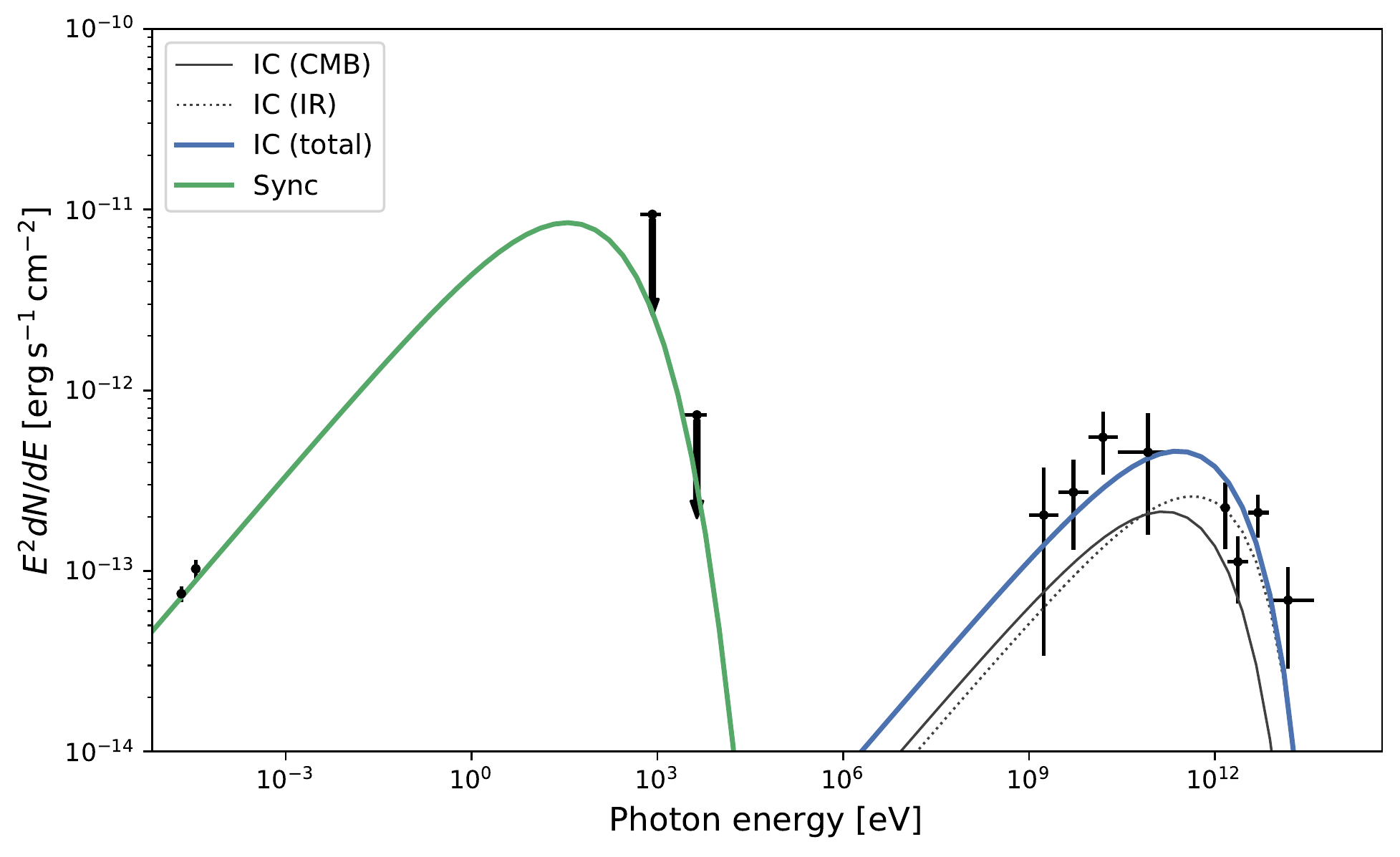}
    & \ \ \ \ \ \ \ \ \
    \includegraphics[angle=0, width=.47\textwidth]{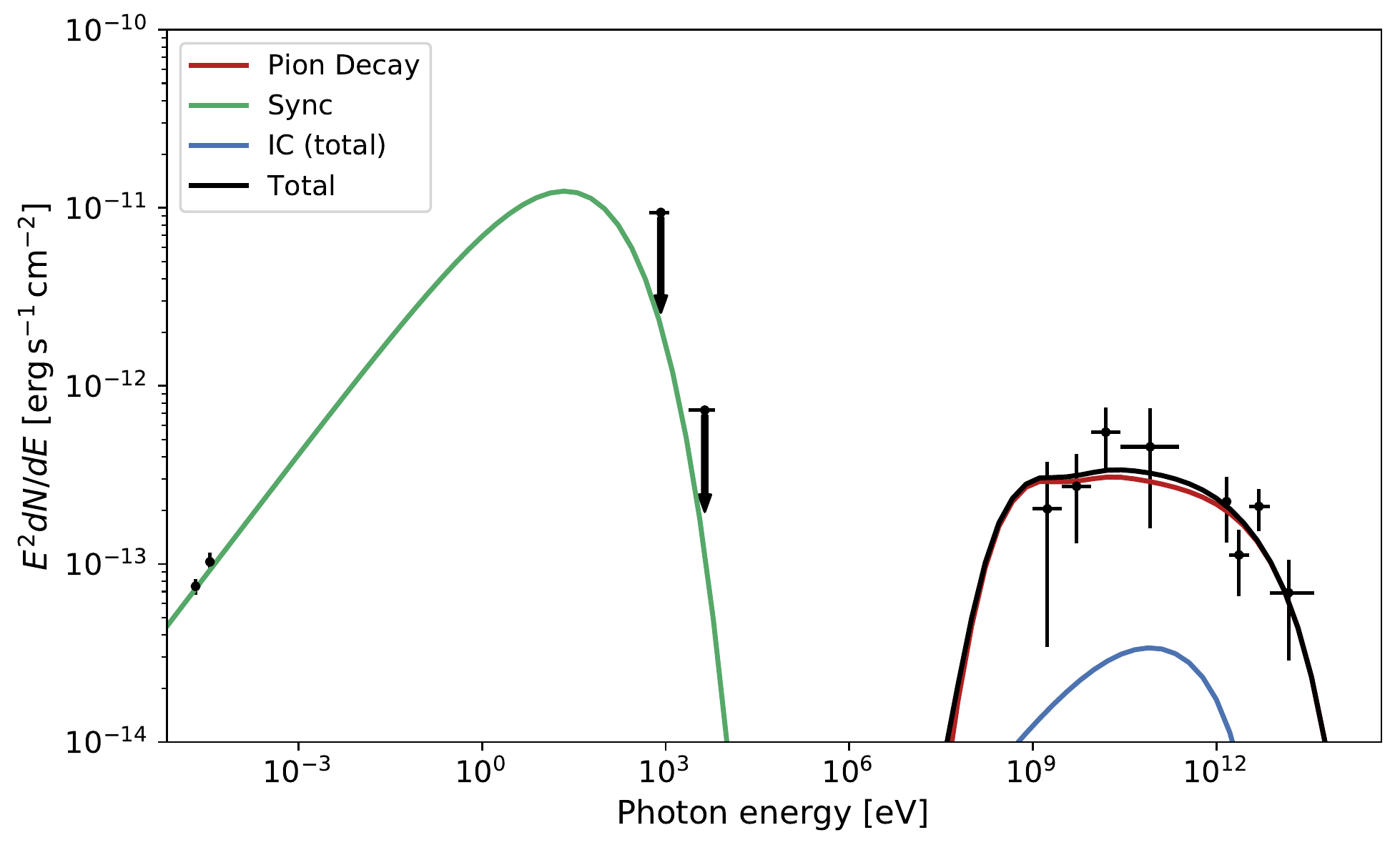} 
    \\
   \end{tabular}
  \caption{Left: Leptonic model considering an electron distribution following a power law with {an} exponential {cutoff}. The parameters are described in Table \ref{tab:fit} and the IR photon density is 
  {1} eV cm$^{-3}$. 
    Right: Hadronic model considering a proton distribution following a power law with {an} exponential {cutoff}. The parameters are described in Table \ref{tab:fit}.} \label{fig:SED_leptonic}
\end{figure*}

The multiwavelength data can be reasonably well modeled  with a purely leptonic model assuming synchrotron emission with a magnetic field of 20 $\mu$G
and inverse Compton emission with two background radiation components, corresponding to the cosmic microwave background (CMB)
radiation with an energy density of 0.26 eV cm$^{-3}$ and an IR component intrinsic to
the remnant itself with a temperature of 145 K and an energy density {of} 1 eV cm$^{-3}$  \citep[see][]{collaboration_exceptionally_2015}.
The latter component is due to dust emission from and around the SNR.
The model assumes an electron distribution following a power law with an exponential cutoff. Fig. \ref{fig:SED_leptonic} shows the model,
and the corresponding parameters are reported in Table \ref{tab:fit}. For this model, the required total electron energy is
W$_{e}$( > 1 GeV) = $4.50\times 10^{49}$ erg, which is {significantly higher than} observed in {other} well-studied SNRs (as discussed in section \ref{sec:ISM}).
Even if an initial released energy higher than the canonical value is considered,
for instance, $\sim$5$\times10^{51}$ erg \citep[see, e.g.,][]{bamba_transition_2018}, then 10 $\%$ of this energy can be transmitted to CR protons,
which {would imply} a maximum electron energy of W$_{e}\simeq 5\times10^{48}$ erg if $n_e/n_p = 0.01$ is assumed.

{Diffusive shock acceleration should lead to an {initial} particle distribution
that is a power law in momentum  \citep[{e.g.,}][]{Malkov2001}. This means that the energy distribution breaks
around the rest-mass energy of the particles. Hence,{
diffusive shock acceleration produces an electron spectrum that} is expected to continue as an {unbroken power law
to energies} well below 1~GeV \citep[e.g.][]{Asvarov_90,Vink_2008}, unlike the hadronic
CR distribution. 
For example, if a lower limit at 10 MeV is considered, then W$_{e}$( > 10 MeV) = $1.3\times 10^{50}$ erg, which suggests that an {implausibly} high
fraction of $>2$\% of the explosion energy would be contained in relativistic electrons.}

The value of 20 $\mathrm{\mu G}$ for the magnetic field strength required by the leptonic model is similar to what was {inferred} by \cite{collaboration_exceptionally_2015} {and} \cite{ bamba_transition_2018}.{ Although some 1000--3000 yr old SNRs do have similar magnetic fields
\citep{bamba05,helder12,zeng19},
such a} magnetic field strength {is surprisingly low for}
a radio source {as luminous} as N132D { ($\sim$38 times more luminous at 1 GHz than the $\sim$1835 yr old SNR RCW 86, which is a VHE $\gamma$-ray source)}, and this would imply that the magnetic field energy density is far out of equipartition with relativistic particle energy density. The equipartition magnetic field strength can be derived from the minimum energy principle, which
minimizes the total energy of the nonthermal
particles plus the magnetic field energy for
a given radio luminosity. The 
magnetic field strength thus derived,
B$_{min}$, is close to the equipartition value \citep[][]{Burbidge1956}.
{It} leads to the following relation for B$_{min}$ \citep[for a review, see][]{Longair:2011,vinkbook}:
{$B_{min}  \mathrm{(G)}= 9.3\times 10^3 \left(\frac{\eta L_{\nu}}{V}\right)^{2/7}\nu^{1/7}$, with $ L_{\nu}$, the radio luminosity
in erg s$^{-1}$ Hz$^{-1}$} at a frequency $\nu$ in Hz, {$V$ corresponds to 25$\%$ of the volume in {cm$^{3}$ }of a sphere with the diameter of N132D (28~pc), and $\eta$ is a parameter taking into account the ratio of the energy present in CR nuclei versus electrons
 (1$<\eta <$100).}
Considering the radio luminosities reported {by} \cite{dickel_radio_1995} {($L_{\nu} =4.3 \times 10^{24}$ erg s$^{-1}$ Hz$^{-1}$ at 5 GHz)} and
the volume corresponding to {a} diameter of 28 pc, {$B_{min}\approx 35\, \mathrm{\mu}$G} is obtained for $\eta = 1$ (the relativistic electron density equals that of CR nuclei), and
134 $\mathrm{\mu}$G {for} $\eta = 100$. 

Fig. \ref{fig:SED_Brem} shows the purely leptonic model with bremsstrahlung in addition {to the} synchrotron and inverse Compton emission
for {the} electron distribution following a power law with {an} exponential cutoff as described above. For the bremsstrahlung component to be
important, the density of the proton targets has to be as high as $n_{\rm p} = 3~ \mathrm{cm^{-3}}$. At this high density, the pion decay
process {will} play a considerable role, which {violates} the assumption of a pure leptonic model.
In a purely leptonic scenario, bremsstrahlung emission cannot explain the data if the proton density remains low.
In the case of a mixed lepto-hadronic model as described in the next section, a hadronic component would require a lower energy for the 
{protons} and electrons, leading to a higher magnetic field value. 
{Then} bremsstrahlung emission is marginal and not needed to explain the data.
It can be concluded here that a purely leptonic model is unrealistic for the {HE-VHE} gamma-ray emission {considering the extreme total energy in electrons required to explain the measured spectra},
and a dominant component of hadronic origin is required.

\begin{table*}[ht]
\centering
\begin{tabular}{cccccccc}
\midrule
MODEL&                  W$_e$(> 1 GeV)   &   $\Gamma$    &           $E_{c_{elec}}$    & B     &W$_p$(> 1 GeV)& n$_p$ &$E_{c_{prot}}$ \\
&erg& & TeV&$\mu$G&erg&cm$^{-3}$&TeV\\
\midrule
Leptonic    &       4.5 $\times$ 10$^{49}$     &   2.2 &           {8}   &   20&--&-- &-- \\

Hadronic    &       4 $\times$ 10$^{48}$            &   2.1    &   2.5  &   100      &4$\times$ 10$^{50}$       &   {10}   &120           \\
\bottomrule
\label{tab:fit}
\end{tabular}
\caption{Parameters for the {proposed} {models} of {the} N132D multiwavelength spectrum. For details, see text. }
\end{table*}

\subsection{Hadronic model}
\label{sec:hadronic}

The {HE-VHE gamma-ray} spectrum can be well fit with a pion decay model, assuming a proton distribution following
a power law with {an} exponential cutoff. The model requires {an ambient} proton density of $n_{\rm p} = 10$ cm$^{-3}$ \citep{hughes_asca_1998},
a total {CR} proton energy  W$_p$(>1 GeV) =  $4 \times 10^{50}$ erg, {a power-law spectral} index  $\Gamma = 2.1 $, and  a cutoff energy of $E_{c}$ =120 TeV
(Table \ref{tab:fit}).
The cutoff energy value is remarkably high: a lower limit is set at $E_{c}$ =45 TeV with 95 $\%$ CL, which is much higher
than the cutoff in the proton spectrum for Cas A, which is estimated to be at $\sim$10 TeV by \cite{CasA_magic:2017}.
The total energy in CR {protons}, $W_p$(>1 GeV) =  $4 \times 10^{50}$ erg, is high but not unrealistic because the energy released in the explosion is thought to exceed 10$^{51}$ erg \citep[e.g.,][]{dickel_radio_1995}, 10$\%$ of which can be {transferred to hadronic {CRs}.} 
The pion decay fit can be {achieved} with a lower {CR} proton energy and a higher {ambient} proton density, for example,
$W_p$(>1 GeV) = 1 $\times 10^{50}$ erg and $n_p= 40$ cm$^{-3}$, but higher proton densities would necessitate additional target material, other than the SNR itself, for the hadronic interaction to take place, for example, a nearby molecular cloud.

Radio and X-ray data were fit with a synchrotron emission model assuming that the electron energy distribution follows a power law with {an}
exponential cutoff, the normalization being constrained by a total electron energy
$W_e$(>1 GeV) = $4 \times 10^{48}$ erg, assuming 1/100 of the proton energy. The model requires a high magnetic field strength {in order to make the leptonic gamma-ray emission subdominant,}
set here to $B= 100~\mu$G, with a cutoff in the electron distribution at E$_{c}(100\mathrm{\mu G})\leqslant$ 3.5 TeV, in order to fit the radio data points and ensure that the X-ray synchrotron component does not exceed the X-ray upper limits. For the spectrum presented in Fig. \ref{fig:SED_leptonic}, E$_{c}$ = 2.5 TeV was chosen (see parameters in Table \ref{tab:fit}). 
{In this model, a cutoff energy in the proton distribution is 
significantly higher than that in the electron distribution.}\footnote{
{In reality, the electron spectrum may even have a break well below 2.5 TeV as a cooling break
is expected at $\sim 0.5$~TeV for $t=2500$~yr and $B=100~{\rm \mu G}$ \citep[see Sect. 13.3][]{vinkbook}}.}

Our model is compatible with similar models in the literature, such as those proposed for Cas A {by} \cite{CasA_VERITAS}, and to some extent, the model {for} {RX J1713.7-3946} \cite[see][]{HESS_RXJ1713_2018}, both of which normalize the electron distribution to the radio and X-ray synchrotron emission.
The chosen magnetic field strength is close to the equipartition value of 135 $\mu$G. The $W_e$/$W_p$ ratio depends on the electron-to-proton ($n_e/n_p$) ratio, but also on the respective volumes for integrating proton and electron densities,
leading to total energy values. CR electrons can fiducially be considered to {emit synchrotron radiation} within the SNR volume.

If {gamma-ray emission} from both CR electrons and protons originates within the SNR, {then the assumption of W$_e$/W$_p$ = $n_e/n_p$ = 1/100 is {reasonable}. However, if gamma-ray emission is mainly produced by escaped CR protons interacting with a molecular cloud, then the W$_{e}$ /W$_{p}$ ratio can have a different value.}

\section{Discussion}
\label{sec:discuss}

The analysis of 252 {hours} of H.E.S.S. observations of N132D confirms the previously published result that N132D is
an exceptionally luminous SNR in the {VHE} band.
The well-determined distance to the LMC makes our luminosity estimation robust{;} it is therefore worthwhile to compare the gamma-ray SED of N132D to the SEDs of the most luminous SNRs in gamma rays. 
{We} found that only one Galactic SNR, G338.3-0.0, is firmly established to show a {VHE} luminosity that is comparable to that of
N132D (see Appendix \ref{SNR} for details).
G338.3-0.0 is {a composite SNR identified with the {VHE} source} HESS J1640-465, {but it is not clear whether
the {VHE} emission comes predominantly from the central PWN or the SNR shell \citep{HGPS_2018}.}  

Another important and new result of the analysis presented here is that no clear cutoff  has been found in the gamma-ray spectrum{, which extends} up to 15 TeV. {The last bin of the
spectrum presented in Fig. \ref{fig:HESS_Flux} corresponds to gamma rays with energies of about 10 TeV or beyond, which is remarkable for such a distant SNR.} 
The spectra of only a few SNR shells reach such a high energy, and they are younger and more nearby sources than N132D. {Assuming a hadronic origin for the} gamma-ray emission from N132D, this implies that protons are present in N132D with energies of at least one hundred TeV.  

The extremely high {VHE} luminosity and the extension of {the} high-energy power-law tail to 15 TeV 
are two observational properties that need to be examined in the context of the physics of SNRs and their environment. 
For this purpose, the {VHE} gamma-ray properties of N132D were compared with {those} of SNRs with similar physical properties, such as age, shock speed, and circumstellar structure (stellar wind bubbles and vicinity of molecular clouds).

\subsection{{Comparison of properties among SNRs}}
\label{sec:SNRs}
\begin{figure}[ht]%
    \centering
 {{\includegraphics[width=0.49\textwidth]{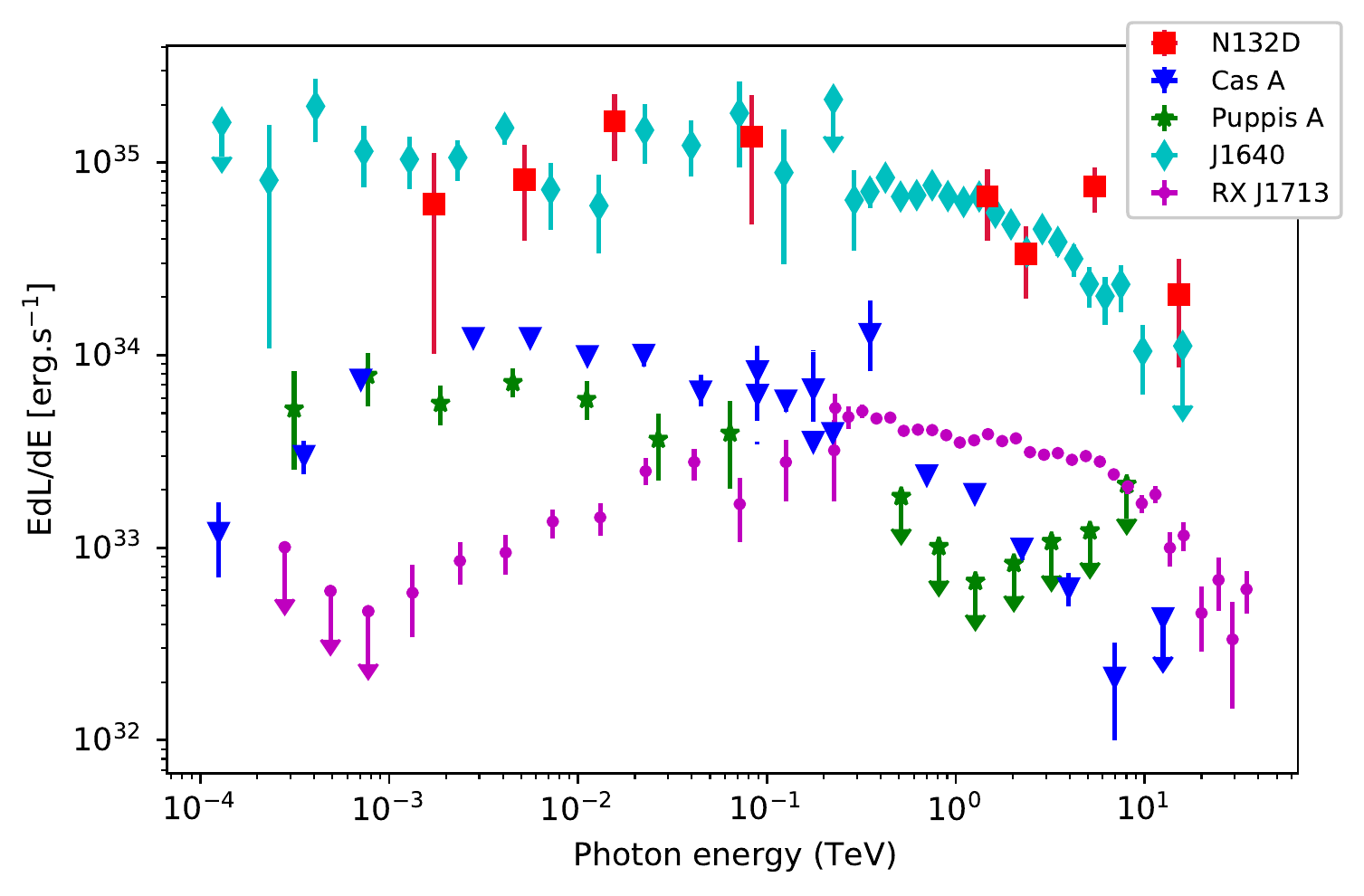} }}%
    \caption{Comparison of differential luminosities, EdL/dE, for {VHE} bright SNRs. Luminosities are computed using the differential flux data points and estimated distances in the following references: N132D, this work, d = 50 kpc; Cas A, \cite{CasA_magic:2017}, $d=3.4$ kpc; 
    Puppis A , {\cite{PuppisA:2015}}, $d=2.2$ kpc; 
    {HESS J1640-465}, \cite{J1640_lemoine-goumard}, $d=10$ kpc; and 
    {RX J1713.7-3946}, { \cite{HESS_RXJ1713_2018}}, $d=1$ kpc.
     {The different shapes of the individual SEDs probably reflect the
     different emission mechanisms (leptonic or hadronic) and evolutionary
     stages. See the original references for detailed spectral models.
     }
    }%
     \label{fig:Luminosities}%
    
\end{figure}

Supernova remnants are {thought} to be efficient accelerators during time intervals when the shock speed is high and they transit from the {ejecta-dominated} phase to the Sedov phase. SNR RX J1713.7-3946 is one of the brightest {VHE}-emitting SNR and is often considered as a prototypical 
CR accelerator. It is younger \citep[1600-2100 yr, e.g.,][]{Tsuji2016} than N132D and it is a nearby source (at 1 kpc). It has a high shock 
speed of $\sim$3500 km s$^{-1}${, therefore} the conditions are 
fulfilled
for rapid diffusive shock acceleration. Due to
its gamma-ray brightness, its spectrum is measured well beyond 10~TeV,
and it exhibits a cutoff at $E_{\mathrm{c}}=12.9 \pm 1.1$ TeV \citep{HESS_RXJ1713_2018}. 
For this source, the leptonic scenario is as plausible as the hadronic one. For a hadronic scenario, the cutoff in the
parent proton distribution is estimated at $E_{c}=88$ TeV, which is similar to our results. With {an} age of about 2500 yr, N132D is about an order of magnitude more luminous in the {VHE} band than SNR RX J1713.7-394{6}. 
Thus, even if the age of SNRs is one of 
the factors making N132D a {source with a high CR energy content,} other factors should also play a role in making this source so luminous. 
The X-ray emission from RX J1713.7-394{6} is dominated by synchrotron radiation from
10-100 TeV electrons, whereas there is no evidence for X-ray synchrotron emission from N132D, whose
X-ray emission is dominated by thermal radiation.
Figure \ref{fig:Luminosities} shows the differential luminosities for five SNRs demonstrating that
N132D is {one of the} two most luminous SNRs, {together} with 
HESS J1640-465. {In all these {VHE} sources, only {the distant but luminous N132D and the nearby RX J1713.7-3946} exhibit flux points above 10 TeV. }

{The gamma-ray emission properties of SNRs have recently been compared by \citet{zeng19} and \citet{suzuki20} using large SNR samples. A comparison of these results to the properties of N132D presented here confirms the exceptional nature of N132D. In particular, the inferred gamma-ray cutoff energy of $E_{c}\gtrsim 8$ TeV is higher than those of the SNRs in Fig. 5b in \citet{suzuki20}, who used a spectral model comparable to the one used by us. Similar conclusion can be drawn from Fig. 3 of \citet{zeng19}, however, care should be taken wherever a lower limit to the cutoff energy is provided for spectra with a broken power-law shape, as this is an additional model complexity.}

N132D belongs to the small class of oxygen-rich SNRs \cite[see][]{Vogt_Dopita_2011}, together with Cas A and Puppis A. These SNRs are characterized
by elevated abundances of oxygen, neon, and other heavy elements, indicating that this material has its origin 
within the helium-burned layers of massive progenitor stars. For the progenitor of N132D, different values {for} its mass have been
suggested, for instance, $>35 M_{\odot}$ \citep[][]{France2009} and $15\pm5\,M_{\odot}$ \citep[][]{Sharda2020}, while the progenitors of 
Cas A and Puppis A have masses in the range of $15-25 M_{\odot}$ \citep[][]{Young2006, Hwang2008}. The comparison of N132D with 
Cas A shows that the two remnants have similar properties in {the radio and X-ray bands}, and bright oxygen emission. 
The distance of Cas A is measured to be 3.4 kpc \citep[][]{CasA_dist_Reed:1995}, and its age is $\simeq$340 years \cite[][]{Thorstensen2001}. N132D is therefore much older than Cas A. However,
Fig. \ref{fig:Luminosities} clearly shows that N132D is much more luminous than its younger sibling. As already mentioned, the
Cas A spectrum exhibits a cutoff at 3.5 TeV that is absent in the N132D spectrum. Puppis A is an older oxygen-rich SNR with 
an age estimated to be $4450\pm750$ years and a  distance of $2.2\pm0.3$ kpc. It is very similar to N132D in terms of chemical composition 
and age, but it is 30 times closer. Puppis A has  not been detected in {VHE} gamma rays \citep[][]{PuppisA:2015}, and its GeV emission is thought 
to be due to the interaction with a nearby molecular {cloud}. A recent study concluded that there is a spectral break at GeV energies, which could explain 
the absence of {VHE} emission \citep[][]{PuppisA_fermi_revisited:2017}. In this regard, Puppis A could be similar to the mature 
SNRs W44 or IC443, whereas, as demonstrated by \citet[][]{bamba_transition_2018}, N132D could be in the transition from being  young to a intermediate-age SNR.
These facts show that there is significant variation in the observational gamma-ray properties {among} the members of the oxygen-rich SNR class. 

\subsection{Interstellar environment}
\label{sec:ISM}

\begin{figure}[ht]
    \centering
   \includegraphics[width=0.49\textwidth]{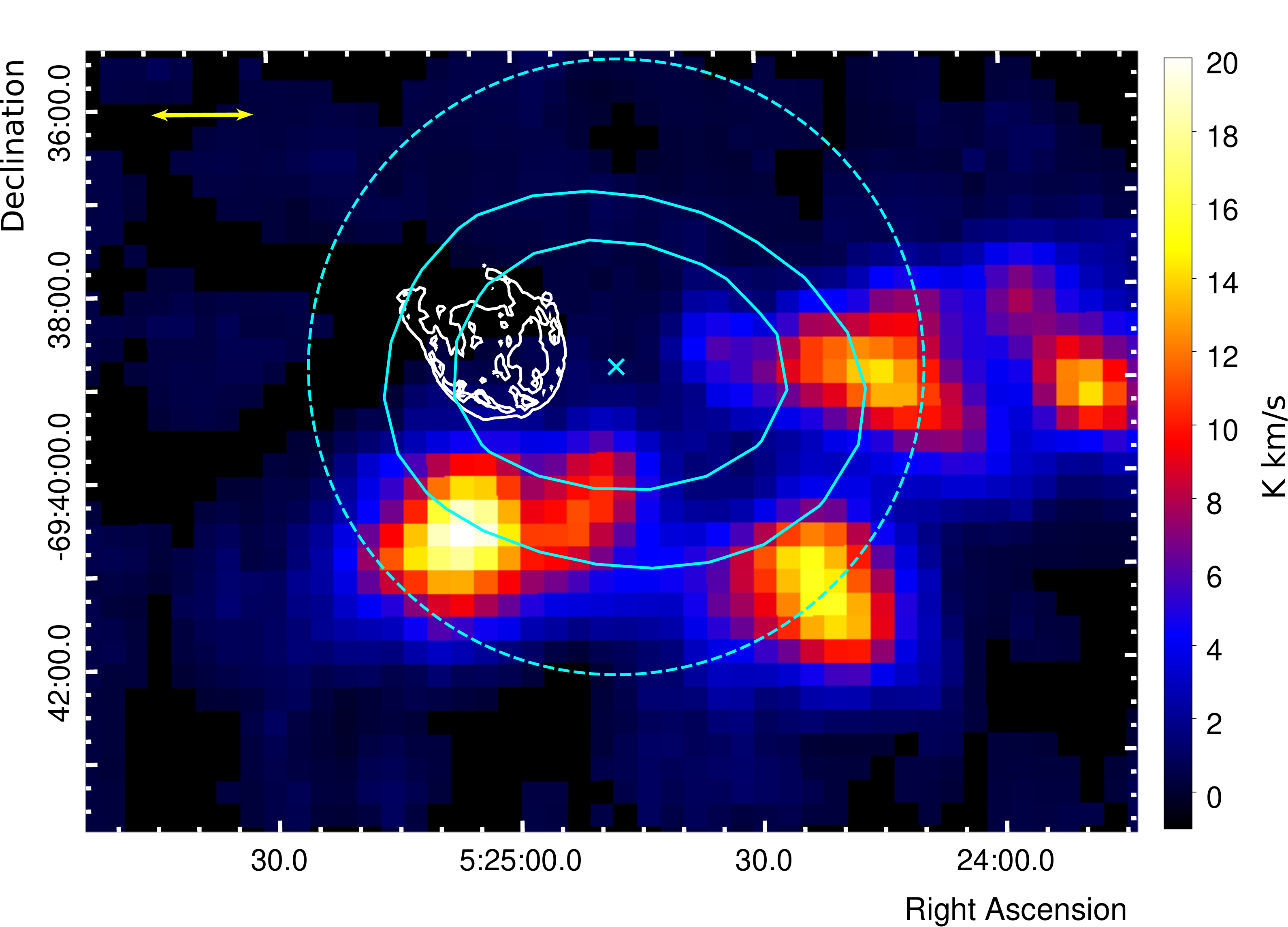}
    \caption[Caption for LOF]{Molecular clouds in the vicinity of N132D. The image shows the CO emission (color scale in K km/s) from {the} MAGMA survey\footnote{https://www.atnf.csiro.au/research/MAGMA/DR1/} {\citep{wong11}}. White contours denote \textit{Chandra} X-ray emission from N132D\footnote{https://chandra.harvard.edu/photo/openFITS/} \citep{borkowski_x-ray-emitting_2007}. The best-fit position of the gamma-ray emission is denoted by the cross, and the {elliptical contours} are the confidence regions of the position {at $2\sigma$ and $3\sigma$}. The dashed circle illustrates the {99\% CL} UL on the extension established at {$3.3^{\prime}$}. The scale bar indicates $1^{\prime}$.} %
    \label{fig:morphology}%
\end{figure}

For hadronic emission, the {VHE} luminosity is determined by the total energy in CRs above $\sim 10$~TeV 
and the local gas density. The high TeV gamma-ray luminosity of N132D may therefore reflect a combination of
a high energy in CRs, perhaps caused by a relatively high explosion energy for this SNR,  
and a high gas density. 
Two possible locations of a dense gas are discussed below, including the shell of a wind-blown
cavity and {a} massive molecular {cloud} near N132D.
Our multiwavelength modeling indeed suggests that the {gamma-ray} emission {most} likely has a hadronic origin. 
{In} the case of a leptonic scenario, the required total energy in the electrons is high compared with other observed SNRs \citep[see also][]{bamba_transition_2018}. {For example, a leptonic model for the gamma-ray emission from  SNRs with a low GeV-to-TeV flux ratio, RX J1713{.7-3946} and {RX J0852.0-4622} \citep[e.g., see Fig. 3 in][]{Dermer_2013}, implies a total electron energy, $W_{\mathrm{e}} (\mathrm{E}>1 \mathrm{GeV})$, of $3.1\times10^{47}$ erg \citep[][]{HESS_RXJ1713_2018} and $2.24\times10^{48}$ erg \citep[][]{Velajr_2018}, respectively.}
For HESS J1640-465, a similar modeling leads to $W_{\mathrm{e}}=10^{49}$ erg \citep[][]{J1640_hess_2014}. 
To a lesser extent, the lack of nonthermal X-rays from N132D is also in favor of a hadronic origin, although nonthermal X-ray emission may be hidden below the 
bright thermal X-ray continuum. {An X-ray synchrotron luminosity similar to that of RX J1713.7-3946 cannot be excluded.}
Moreover, the radio luminosity of N132D suggests a high magnetic field strength, 
compatible with  the {100 $\mu$G strength {assumed} in the hadronic model presented in section \ref{sec:hadronic}}.

For the preferred hadronic scenario for the gamma-ray emission of N132D, a high gas density is required, and the question
arises where most of these gamma rays are coming from:
from CRs within the densest portions of the SNR shell, or from CRs that have escaped the shell and are interacting 
with the massive molecular cloud near N132D \citep{Banas_1997,Sano:N132D}{.}  
While the GeV
luminosities of SNRs interacting with molecular {clouds} are higher than those of SNRs without nearby molecular clouds,
this tendency does not hold for {VHE} gamma rays \citep[e.g.,][]{Fernandez2013}. 
The {VHE} luminosity enhancement for N132D and
its hard gamma-ray spectrum at {VHE} energies can be interpreted under the assumptions that (1) the shock accelerates
protons until the highest possible energy is reached, and at this point, the shock can no longer confine the protons,
and (2) the most energetic protons escape {then} into the interstellar medium (ISM). The gamma-ray emission created by the interactions of these
escaped protons with a gas {cloud} will be strong and hard \citep[][]{Aharonian_1996,Gabici2007, Moskalenko2008}. 
The hadronic scenario requires a high density of the target material.
These high densities can be found in molecular clouds. SNRs such as W28 or W44 show gamma-ray emission originating in
nearby molecular clouds \citep[see, e.g.,][]{hess_collaboration_W282008,cardillo_supernova_2014,makino_interaction_2019}, indicating that protons accelerated in the SNR bombard the molecular cloud. N132D is interacting with a molecular
cloud as well, which has recently been shown {by} \cite{Sano:N132D}. Furthermore, the X-ray
emission is brightest in the southwestern part,
where the SNR is interacting with dense matter. 

The molecular {clouds} near N132D {as traced by the CO emission} {are} shown in Fig.~\ref{fig:morphology}. {The velocity of the clouds, $+264$ km s$^{-1}$, agrees with that of N132D, $+268$ km s$^{-1}$} \citep[][]{Banas_1997}. The {best-fit}
position of the gamma-ray emission ({elliptical contours} in Fig.~\ref{fig:morphology}) is
consistent with the position of the SNR itself and with parts of the
molecular cloud. Therefore these two regions are potential sites of {gamma-ray emission}. {If the {VHE} gamma-ray emission comes from these molecular clouds, a diffusion coefficient of $D\simeq2\times10^{29}\left(t/2500\ \ \mathrm{yr}\right)^{-1}$ cm$^{2}$ s$^{-1}$ for multi-TeV CRs is required.} To determine the spatial origin of the gamma-ray emission, a higher sensitivity and a finer angular resolution are needed. 
Nevertheless, the spectrum with an unbroken power law from GeV to TeV energies could be taken as evidence for a single population of emitting particles. This would be difficult to reconcile with illumination of MCs by CRs originating from the shell.
In that case, the low-energy gamma rays would likely come from the shell because only the highest-energy CRs would penetrate the MC and cause gamma-ray radiation at VHE. {In addition to the molecular clouds shown in Fig.~\ref{fig:morphology},
\citet{Sano:N132D} reported smaller clouds that interact
with the SNR, as indicated by the ratio of CO $J=3-2$ versus $J=1-0$ line emission.
These cloudlets are situated in the south and near the center of the SNR. They might also be locations of enhanced
hadronic emission, but then directly associated with the SNR itself.
}

A hadronic scenario involving a wind-blown cavity is also possible. As suggested by \citet[][]{chen_supernova_2003}, 
in case  of a very massive progenitor, a low-density bubble is formed \citep[see also][]{Weaver1977}. When the SN occurs, the shock propagates with 
a high velocity {($v \sim$ 5000-10000 km s$^{-1}$)} into the bubble. When the shock wave reaches the high-density shell of 
swept-up material, it heats the dense shell, and the SNR appears most luminous, but the shock velocity rapidly {decreases}. A wind-blown bubble model was considered for G338.3-0.0 
in order to explain the high {VHE} luminosity of HESS J1640-465 \citep[][]{J1640_hess_2014}. 
Under the assumptions that the SNR shock freely 
expands up to the radius of the wind-blown bubble, and with average shock {speeds}  between 5000 km s$^{-1}$ and 10000 km s$^{-1}$, the authors 
proposed that the age of SNR G338.3-0.0 could be 1000-2000 yr in this case, and at least younger than 5000-8000 yr \citep[][]{Slane2010}. 

The wind-blown bubble model also fits the properties of N132D, given the inconsistency between its kinematic age (2500 yr) and its Sedov dynamical age (6000 yr),
indicating that the supernova explosion occurred within a cavity in the interstellar medium \citep[][]{Hughes1987}. 
In addition to the {VHE} luminosity enhancement, it is conceivable that this model can also explain the presence of sub-PeV 
protons to some extent because an SNR expanding into a preexisting stellar wind can accelerate protons to a higher energy than SNRs expanding in 
a uniform circumstellar medium \citep[][]{Voelk1988}. Because the gamma-ray spectrum of G338.3-0.0/HESS J1640-465 has an exponential cutoff at 
6 TeV, it is unclear so far whether the luminosity enhancement and the increase in a maximum CR energy are expected to be observed simultaneously. 

{Two} other SNRs in the LMC, N63A and N49B, have also been proposed to have evolved 
in wind-blown 
cavities \citep[][]{hughes_asca_1998}. Because N63A and N49B were detected with \textit{Fermi}-LAT \citep[][]{Campana2018} and have massive
progenitors, it is possible that the luminosity enhancement takes place in these SNRs as well, as suggested, for example, in \cite{Sano_ALMA_LMC:2017}. 
This hypothesis can be tested {by} observations of N63A and N49B, which are located in the northern part of the LMC disk about 4$^{\circ}$ away from N132D. 

{In general, the LMC is well suited for {studies} of SNRs \citep[see][]{Ginzburg_72} because its projection is almost face-on (inclination of 35$^{\circ}$) and the distance to the LMC SNRs is more certain than for some Galactic SNRs, allowing us to accurately derive the {VHE} luminosity. Galactic PeVatron candidates with hard ($\Gamma \sim$ 2) and featureless spectra similar to those derived for N132D {in this work} can be listed among unresolved H.E.S.S. sources in the Galactic plane \citep[e.g.,][]{HGSP_PeV}. For example, SNR G318.2+0.1, one of the two most luminous Galactic SNRs {at VHE,} was proposed to be associated with HESS J1457-593 in 2010 \citep{G318_VHE} and is listed among PeVatron candidates. However, the association of Galactic VHE sources is often complicated by the presence of several possible source candidates for a given region within the Galactic plane.

{Because of the lack of evidence} for an energy cutoff in the N132D spectrum, N132D remains a PeVatron candidate.
\cite{Cristofari_2020} {suggested} that only the remnants of very powerful rare supernova explosions such as 
the one that produced the N132D remnant can accelerate CR particles to PeV energies}.
Moreover, the LMC has a significant angular size, allowing 
a detailed study of individual SNRs resolved in the VHE band. {Identifying the origin of CR hadrons is a long-standing problem} \citep{Ginzburg_72}, and the establishment of discrete accelerators of CR protons in the LMC, such as N132D, shows that SNRs are viable sources of CRs {in}  the Milky Way and the LMC.

{The possible role of molecular
cloud interaction on the gamma-ray emission is worth investigating with the
next generation of IACTs, in particular, with the Cherenkov Telescope Array} \citep[CTA;][]{2019CTAbook},
which will have a sensitivity that is an order of magnitude better than H.E.S.S. and its angular resolution will  be higher. 
{Future N132D observations with CTA might therefore establish whether there is a shift between the SNR center and the origin of gamma-ray emission}, and moreover, they might help to determine the energy of the
spectral cutoff or even establish this enigmatic SNR as a CR PeVatron.

\section{Conclusion}
\label{sec:conclusion}

The LMC SNR N132D {is detected} with a statistical significance of 5.7 $\sigma$ {above 1.3 TeV} on the basis of H.E.S.S. observations with an exposure time of 252 hours, which is {104} hours longer than the
exposure used in the previous publication \citep[][]{collaboration_exceptionally_2015}. The inclusion of new observations results in {an unambiguous} detection and allowed us to perform a more detailed spectral analysis. 

The Fermi-LAT and H.E.S.S. gamma-ray spectrum extends up to 15 TeV and is {well described} with a power-law index of {2.13} $\pm$ 0.05. No cutoff in energy is needed to explain the spectrum, as a power law with {an} exponential cutoff fit is statistically as valid as a simple power-law model. 
A 95$\%$ CL lower limit on an exponential cutoff is derived at 8 TeV. This lower limit exceeds the cutoff value of 3.5 TeV established for the spectrum of {$\sim$340-year-old} 
Cas A by \cite{CasA_magic:2017}, although N132D is $\sim$2500 years old. N132D is the only extragalactic SNR detected in gamma rays so far, and its luminosity {is} compatible with that of the most luminous Galactic SNR G338.3-0.0. The absence of a clear spectral break and/or cutoff energy and its high gamma-ray luminosity make N132D a very special object. It is one of the oldest and most distant {SNRs emitting at VHE}.

A purely leptonic model fails to satisfactorily explain the multiwavelength spectrum of N132D. 
The main argument is that the required total energy of electrons is too high 
{compared to the fraction of explosion energy that is expected to be transferred to leptonic CRs.}
An additional but less compelling argument is that the magnetic field strength in the case of a purely leptonic scenario is {required to be}  surprisingly low, 20 $\mu$G, which is at odds with the radio brightness of this SNR.  Finally, the cutoff energy required for the electron distribution is quite low (8 TeV) compared to the absence of a cutoff observed in the gamma-ray spectrum. Our conclusion is that gamma-ray emission from N132D is most likely hadronic in origin.

The hadronic origin of the {VHE} gamma-ray emission raises the question whether its high luminosity
is due to a very efficient acceleration at the SNR shock front, perhaps related to the
evolution of the shock inside a wind-blown cavity,
or to an interaction of the SNR with the molecular cloud, which could enhance its gamma-ray luminosity. A massive molecular cloud is present in the southwest of the remnant, where enhanced emission in radio, optical, and X-ray bands {is} observed. {These two scenarios are possible given the uncertainties in VHE source position}.
Even if some of the emission is caused by CR interaction with the molecular cloud,
the {hard} power-law spectrum, together with the lack of evidence for a {cutoff} energy,
makes the gamma-ray SNR very distinct from mature SNRs that are known to interact
with molecular clouds, such as IC 443, W44, and W28. 

\section*{Acknowledgements} 

The support of the Namibian authorities and of the University of Namibia in facilitating the construction and operation of H.E.S.S. is gratefully acknowledged, as is the support by the German Ministry for Education and Research (BMBF), the Max Planck Society, the German Research Foundation (DFG), the Helmholtz Association, the Alexander von Humboldt Foundation, the French Ministry of Higher Education, Research and Innovation, the Centre National de la Recherche Scientifique (CNRS/IN2P3 and CNRS/INSU), the Commissariat à l'énergie atomique et aux énergies alternatives (CEA), the U.K. Science and Technology Facilities Council (STFC), the Knut and Alice Wallenberg Foundation, the National Science Centre, Poland grant no.2016/22/M/ST9/00382, the South African Department of Science and Technology and National Research Foundation, the University of Namibia, the National Commission on Research, Science $\&$ Technology of Namibia (NCRST), the Austrian Federal Ministry of Education, Science and Research and the Austrian Science Fund (FWF), the Australian Research Council (ARC), the Japan Society for the Promotion of Science and by the University of Amsterdam. 
We appreciate the excellent work of the technical support staff in Berlin, Zeuthen, Heidelberg, Palaiseau, Paris, Saclay, Tübingen, and in Namibia in the construction and operation of the equipment. This work benefitted from services provided by the H.E.S.S. Virtual Organisation, supported by the national resource providers of the EGI Federation. 
J. Vink and D. Prokhorov are partially supported by funding from the European Union’s Horizon 2020 research and innovation programme under grant agreement No 101004131 and by the Netherlands Research School for Astronomy (NOVA).
\appendix

\section{Supplementary material for the multiwavelength study}

\begin{figure}[ht]
    \centering
 {{\includegraphics[width=0.49\textwidth]{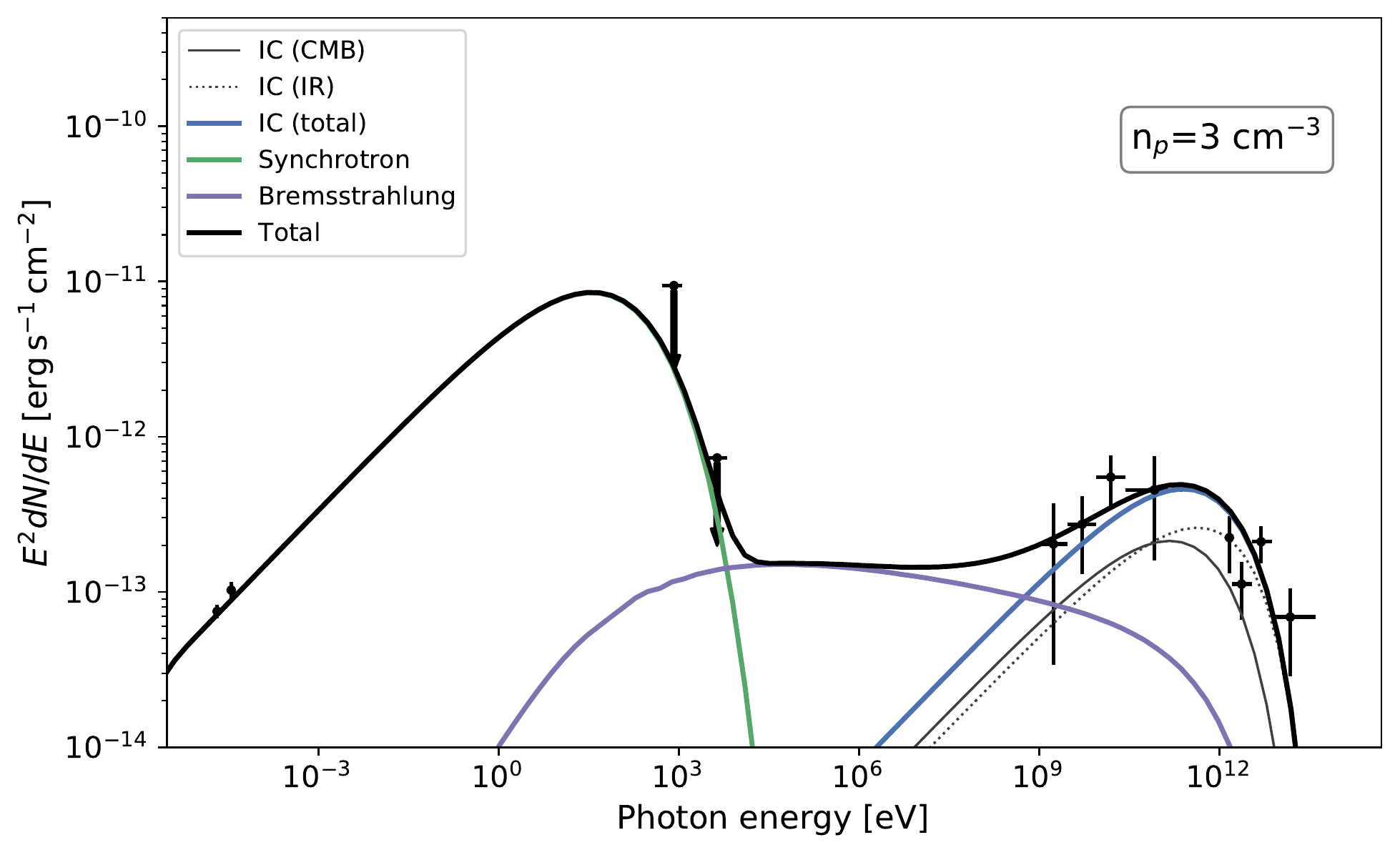} }}%
    \caption{Leptonic model including bremsstrahlung emission. The model considered an electron distribution following a power law with {an} exponential cutoff. The parameters are described in Table \ref{tab:fit}. } %
     \label{fig:SED_Brem}%
\end{figure}
\section{Galactic SNRs}
\label{SNR}
The list of luminous (above 20\% of the {VHE} luminosity of N132D) Galactic SNRs available in the literature includes CTB 37A, SNR G318.2{+0.1}, CTB 37B, SNR G359.1-0.5, SNR G106.3{+2.7}, SNR G8.7-0.1, and SNR {G}338.3-0.0, which are sorted {here} with increasing {VHE} luminosity. While this list consists of seven SNRs, there are counter-arguments to the high {VHE} luminosities of five of these objects. (i) The uncertainties in a background model are suggestive of a lower flux of CTB 37A (which are illustrated in Fig. 5 of \cite{HGPS_2018}) than that published in \cite{CTB37A_2008}, (ii) the estimated energies of the relativistic particles are too high for the hadronic and leptonic scenarios applied to CTB 37B, suggesting that the distance of 13.2 kpc to this source is an overestimate \citep{CTB37B_2016}, (iii) SNR G359.1-0.5 is likely only to be responsible for a part of the VHE gamma-ray source, HESS J1745-303, given that the solid angle of region A \citep{J1745_2008} that embeds a molecular cloud, as viewed from the center of G359.1-0.5, is about an order of magnitude larger than the angles of the other emitting regions \citep[also][]{J1745_2012}, but the contribution from region A to the total flux of HESS J1745-303 is subdominant \citep{J1745_2008}, (iv) the distance to SNR G106.3{+2.7} is uncertain, and if a distance of 0.8 kpc instead of 12 kpc is adopted, then its estimated {VHE} luminosity decreases significantly \citep{G106_2009}, (v) HESS 1804-216 is likely to be dominated by a {PWN} surrounding the pulsar PSR {B1800-21}, but not by SNR G8.7-0.1, as demonstrated through the \textit{Fermi}-LAT spectral and spatial analyses \citep{W30_2019} and on the basis of a comparison of its properties with those of the established {VHE} PWN \citep{PWNpop_HESS2018}.
Finally, SNR G318.2{+0.1} is associated with the {VHE} source HESS J1457-593 and is suggested to be originated from an interaction between G318.2{+0.1} and a molecular cloud seen in $^{12}$CO data 
\citep[][]{Hofverberg2010}, and it therefore cannot be directly compared to SNR N132D. {It can also be noted here} that in the case of the {composite} SNR {G338.3-0.0} associated with
HESS J1640-465, the {scenario in which the {VHE} emission originates predominantly from the PWN}
of the subsequently discovered PSR J1640-4631 is highly rated in the context of the
{VHE} PWN population \citep{PWNpop_HESS2018}; it remains plausible nonetheless that a significant part of the gamma-ray
emission from HESS J1640-465 originates in the SNR shell \citep[][]{J1640_hess_2014}.

%
%

\bibliographystyle{aa}
\bibliography{N132D_References}

\end{document}